\documentclass[
 reprint,
 superscriptaddress,
 amsmath,amssymb,
 aps,
pra,
showkeys
]{revtex4-2}

\usepackage[hidelinks]{hyperref}
\usepackage{comment}
\usepackage{soul} 

\usepackage{physics}

\usepackage{times}

\usepackage{graphicx}
\usepackage{caption}
\usepackage{subcaption}

\usepackage{xparse}
\usepackage{mathtools}
\usepackage{setspace}
\usepackage{tikz}
\usepackage{lineno}
\usepackage[utf8]{inputenc}
\usepackage[T1]{fontenc}
\usepackage{mathptmx}
\usepackage{etoolbox}

\makeatletter
\def\@email#1#2{%
 \endgroup
 \patchcmd{\titleblock@produce}
  {\frontmatter@RRAPformat}
  {\frontmatter@RRAPformat{\produce@RRAP{*#1\href{mailto:#2}{#2}}}\frontmatter@RRAPformat}
  {}{}
}%
\makeatother

\usepackage{caption}
\begin{document}

\title{Modeling the performance and bandwidth of single-atom adiabatic quantum memories}

\author{Takla Nateeboon}
\affiliation{Optical and Quantum Physics Laboratory, Department of Physics, Faculty of Science, Mahidol University, Bangkok 10400, Thailand}

\author{Chanaprom Cholsuk}
\affiliation{Department of Computer Engineering, TUM School of Computation, Information and Technology, Technical University of Munich, 80333 Munich, Germany}
\affiliation{Abbe Center of Photonics, Institute of Applied Physics, Friedrich Schiller University Jena, 07745 Jena, Germany}

\author{Tobias Vogl}
\affiliation{Department of Computer Engineering, TUM School of Computation, Information and Technology, Technical University of Munich, 80333 Munich, Germany}
\affiliation{Abbe Center of Photonics, Institute of Applied Physics, Friedrich Schiller University Jena, 07745 Jena, Germany}

\author{Sujin Suwanna}
\email{sujin.suw@mahidol.ac.th}
\affiliation{Optical and Quantum Physics Laboratory, Department of Physics, Faculty of Science, Mahidol University, Bangkok 10400, Thailand}

\date{\today}

\begin{abstract}
Quantum memories are essential for quantum repeaters that will form the backbone of the future quantum internet. Such memory can capture a signal state for a controllable amount of time after which this state can be retrieved. In this work, we theoretically investigated how atomic material and engineering parameters affect the performance and bandwidth of a quantum memory. We have applied a theoretical model for quantum memory operation based on the Lindblad master equation and adiabatic quantum state manipulation. The materials’ properties and their uncertainty are evaluated to determine the performance of Raman-type quantum memories by showcasing two defects in two-dimensional hexagonal boron nitride (hBN). We have derived a scheme to calculate the signal bandwidth based on the material parameters as well as the maximum efficiency that can be realized. The bandwidth depends on four factors: the signal photon frequency, the dipole transition moments in the electronic structure, cavity volume, and the strength of the external control electric field. As our scheme is general and independent of materials, it can be applied to many other quantum materials with a suitable three-level structure. We therefore provided a promising route for designing and selecting materials for quantum memories. Our work is therefore an important step toward the realization of a large-scale quantum network.
\end{abstract}

\keywords{quantum memory, quantum repeater, defects, hexagonal boron nitride, quantum network, quantum communication}

\maketitle
\section{Introduction}
Quantum communication is an emerging technology that promises unconditional security for encryption enabled by fundamental physical principles \cite{PhysRevLett.67.661, BENNETT20147, science.283.5410.2050}. A global network connecting many users termed the quantum internet has been proposed for many years \cite{10.1145/1039111.1039118, Kimble2008, vanmeterbook2014, Pirandola2016, Simon2017, Sasaki2017, science.aam9288, Khatri2021}. Quantum teleportation over short distances up to several hundreds of kilometers through fibers has been achieved \cite{Brekenfeld2020, Rabbie2022, notzel2022operating, Pittaluga2021}. Similarly, free-space quantum communication between two distant nodes had been demonstrated using fibers and satellites \cite{science.aan3211, Liao2017, Chen2021, deForgesdeParny2023}. For long-distance communication and distributed quantum computing, a quantum network is required which will likely be of hybrid nature composed of different quantum systems, free-space and fiber links, as well as different protocols and components \cite{science.aam9288, 10.1117/12.2574154, Matsuo2019, khatri2021policies, pathumsoot2021optimizing, Satoh2021, Abasifard2023, quick3}. To build such network, multiple devices must be integrated together efficiently. One essential building block is a quantum repeater which enables entanglement swapping via the Bell state measurement (BSM) \cite{PhysRevLett.70.1895, PhysRevLett.80.1121, Duan2001, gottesman2012longer, Lee2013BellstateMA, PhysRevLett.81.5932}. In order for a quantum network to transmit and process quantum information efficiently, however, the signals between distant nodes need to be synchronized which poses a great challenge for the currently available state-of-the-art technology \cite{8910635,Bhaskar2020, Wang2021}.\\ 
\begin{figure}
    \centering
    \includegraphics[width = 0.46\textwidth]{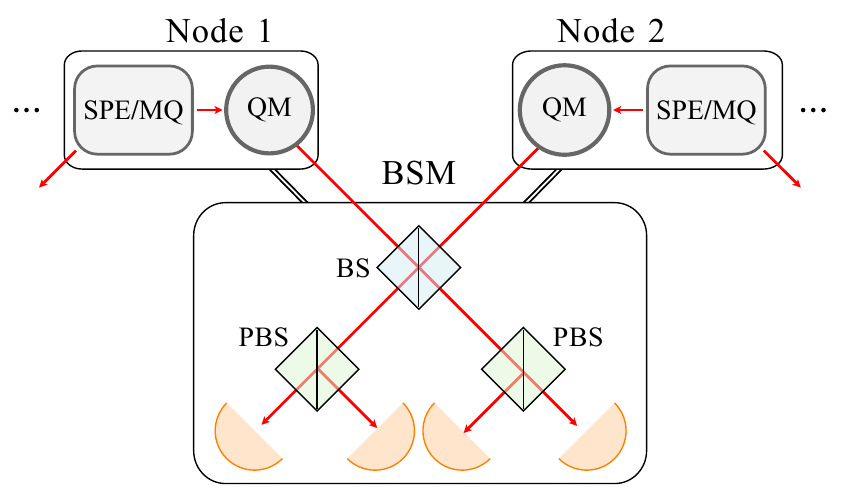}
    \caption{
        A schematic overview illustrates the roles of single photon emitter (SPE) and quantum memory (QM) in a quantum network. Here, MQ denotes matter qubits that can be entangled with SPE, which produces photonic qubits. A quantum memory can synchronize two photonic qubits in a Bell state measurement (BSM) for entanglement swapping. A classical communication channel between the entanglement swapping unit and each node is denoted by double lines. The schematic is inspired by the concepts in Refs.~\cite{Hartmann2007,PhysRevA.71.050302}.} 
        \label{fig:schematic memory}
\end{figure}
\indent By the no-cloning theorem, a quantum state cannot be arbitrarily copied. Hence, one approach to repeat quantum states and synchronize two quantum nodes is by utilizing a quantum memory \cite{Hartmann2007, Lvovsky2009, Simon_2010, sciadv.abn3919, Heshami2016, li2019experimental,Guendogan2021}, which can be thought of as a device that couples a photonic state and a matter state (also termed a matter qubit here) as depicted in Fig.~\ref{fig:schematic memory}. In principle, a quantum memory will be activated by an incoming photon and transferring a photonic state to a matter qubit, which will be kept for a certain duration. When a photon is needed again, a \emph{reverse} procedure is applied to retrieve the photonic qubit from the matter qubit. With quantum memory units, a quantum network can synchronize the signal in a controllable and flexible manner. In addition, quantum computers will require a quantum memory as well - either to store quantum information during an algorithm while other parts of the logic are performed or to link and synchronize distant quantum computers. In these cases, the quantum memory would be the quantum analog of the classical random access memory (RAM).\\
\indent There are many types and physical realizations of quantum memories, each with its own advantages and disadvantages. For instance, the adiabatic type of quantum memory has the desirable property of a high efficiency \cite{Simon_2010, Cho2016}. This type of memory can be realized by an atomic gas vapour, e.g. Rubidium or a solid-state system \cite{Poem2015}. The vapour cell memory suffers from four-wave mixing and Doppler broadening \cite{PhysRevA.87.063836, Offer2018, Camacho2009}. On the other hand, while Doppler broadening can be mitigated in solid-state systems \cite{NILSSON2005393}, their phonon coupling \cite{Gorshkov2007-3, Heshami2016, Zhu2011, Khalid2015} and spectral diffusion \cite{PhysRevLett.116.033603, Bradac2010} remain excessively large.\\
\indent One of the solid-state systems realized as a quantum memory is the well-known nitrogen-vacancy center in diamond (NV center), which demonstrated storage time in the order of milliseconds by exploiting their nuclear spins \cite{Guo2019, Heshami2014, PhysRevA.97.052303, NV-naturephy}. In addition, this material can act as a single photon emitter and a quantum sensing device \cite{Nemoto2016}; however, its efficiency is commonly limited by high internal and Fresnel reflection which leads to high coupling losses, alongside strong phonon modes \cite{Wan2018}. Another prominent candidate is silicon vacancy center in diamond \cite{Berezhnoi_2022}, which has a low radiative lifetime \cite{Bradac2019} and has been shown to perform with a high fidelity as a quantum memory \cite{PhysRevLett.119.223602}. Alternatively, two-dimensional (2D) materials like hexagonal boron nitride (hBN) might be more promising due to near-ideal outcoupling efficiency \cite{C9NR04269E}, robustness in harsh environments \cite{Vogl2019-radiation,doi:10.1021/acsphotonics.7b00086}, relatively low phonon coupling \cite{10.1021/acsphotonics.8b00127}, as well as a large number of defect choices \cite{ Cholsuk2022,cholsuk2023comprehensive}. These advantages make hBN defects potentially viable candidates for quantum memory.\\
\indent Regardless of its origin or base physical system, a quantum memory must be evaluated by the performance metrics, which include efficiency, storage time, fidelity, operable bandwidth and integrability \cite{Guendogan2021, Simon_2010, Lvovsky2009, Simon_2010, Heshami2016}. This performance depends on many factors such as material properties from its electronic structure, resonant frequency, control frequency, and operation time, among others. In this work, we investigated the performance of an adiabatic-type quantum memory, focusing on a single-atom memory unit in a solid-state system and its dependence on material properties (level structure and population lifetimes) and control parameters (detuning, pulse strength, and duration). 
The model is based on the Lindblad master equation with decay and dephasing rates taken into account. Furthermore, we applied the model to analyze the performances of solid-state systems that have required electronic structure to become a quantum memory. This work selected titanium bi-vacancy (Ti$_\text{VV}$) and molybdenum bi-vacancy (Mo$_\text{VV}$) defects in hBN as case studies. Numerical values of fidelity, efficiency, and bandwidth of such defects are evaluated. Both defects are subsequently determined whether they would be feasibly enhanced with a suitable cavity.\\
\indent Compared to the work in Ref. \cite{Specht2011-Nature09997} that focused on single-atom quantum memories, we studied specific parameters of coupling constant $g$, control field strength, channel decay rates 
 $\gamma$, and cavity decay rate $\kappa$ in the units relative to the coupling constant $g$. This allows us to explore the parameters space and identify the relations for evaluating a prerequisite structure of a quantum memory that is independent of a specific physical system.
\section{Framework for evaluating material properties for quantum memories}
\label{framework}
\begin{figure}[ht]

\includegraphics[width = .43\textwidth]{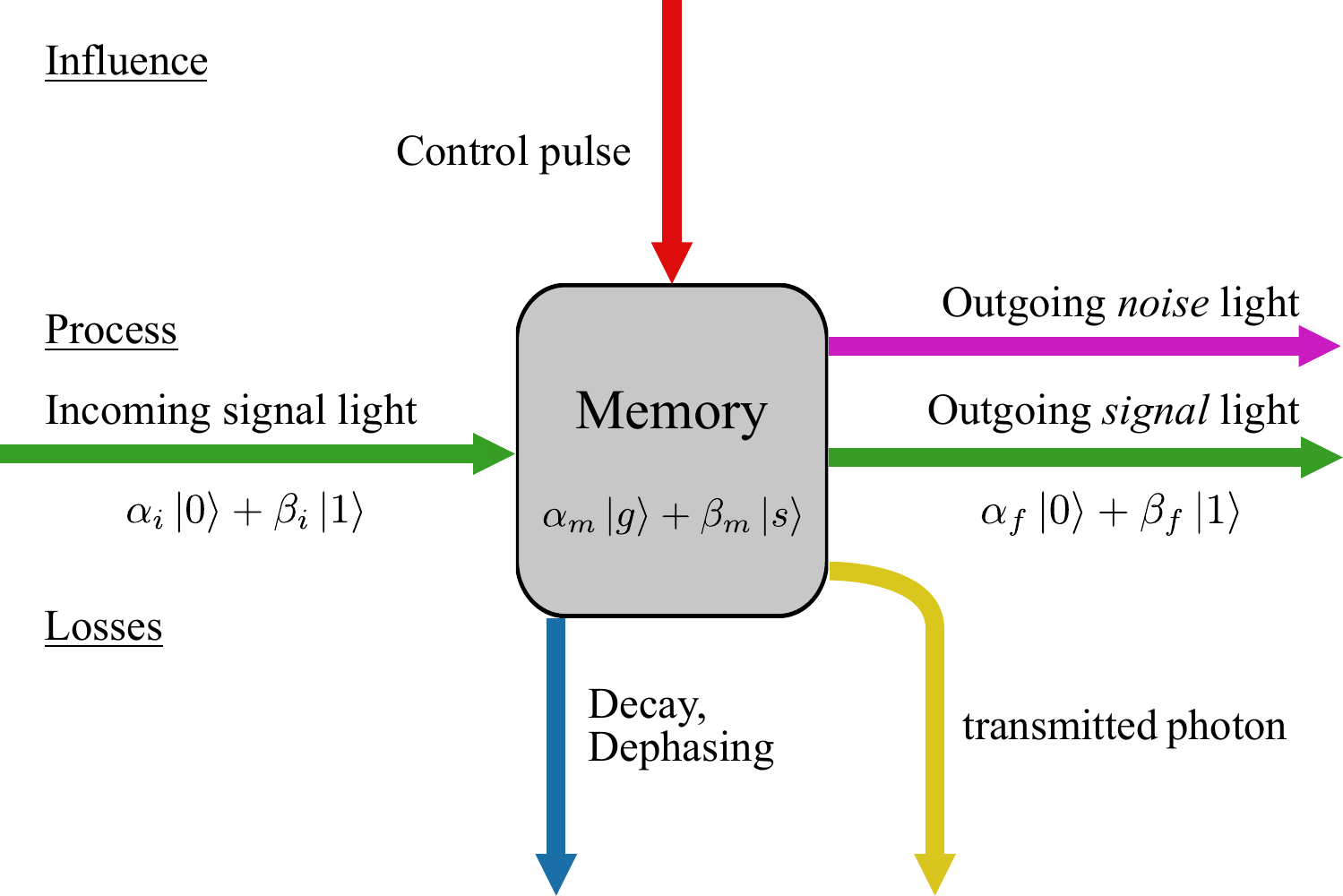}
\caption{Schematic of a quantum memory illustrating all processes, influences, and losses.}
\label{fig:DiagramNoises}
\end{figure}
We evaluated the quantum memory performance based on the dynamical model of a $\Lambda$-type three-level system. We assumed that the memory is in a cavity of volume $V$ containing one signal photon of frequency $\omega_1$. The signal field has a constant coupling $g$ for the entire operation while the control pulse $\Omega(t)$ is of sigmoid shape with characteristic time $T$; that is,
\begin{equation}
    \Omega(t) = \frac{\Omega_0}{1+\exp\frac{t}{T}},
    \label{eq:control-pulse}
\end{equation}
where $T$ is a characteristic time scale for the control pulse and $\Omega_0$ is a peaked Rabi frequency. This shape is for a writing process; whereas the reading process is assumed to be a reverse process.\\
\indent A schematic of the quantum memory flow is depicted in Fig.~\ref{fig:DiagramNoises}. In our work, we focused on the control pulse shape, dephasing rates, and decay rates, while setting the incoming signal photon to be a photon number state $\ket{n} = \ket{1}$ as a controlled variable. To search for quantum memory candidates, we evaluated how efficient the writing process in a quantum memory would be with different combinations of control parameters and evaluate the bandwidth of the quantum memory.

\subsection{Three-level system}
\begin{figure}[ht]
    \centering
    \begin{tikzpicture}
        \draw (1,2.7) -- (3,2.7) node[right] {$\ket{e}$};
        \draw (2,1) -- (4,1) node[right] {$\ket{s}$};
        \draw (0,0) -- (2,0) node[right] {$\ket{g}$};;
        \draw[<->] (1,0) -- (1.9,3.2);
        \draw[dashed] (1,3.2) -- (3,3.2);
        \draw[dashed] (1,3.5) -- (3,3.5);
        \draw[<->] (1,2.7) -- (1,3.2);
        \draw[<->] (3,3.2) -- (3,3.5);
        \draw (3,3.3) node[right] {$\delta$};
        \draw[<->] (3,1) -- (2.1,3.5);
        \draw (1,1.5) node[left] {$g$};
        \draw (2.7,2) node[right] {$\Omega(t)$};
        \draw (0.9,3.15) node[left] {$\Delta$};
    \end{tikzpicture} 
    \caption{Energy level diagram of a $\Lambda$-type system with ground state $\ket{g}$, excited state $\ket{e}$, and metastable state $\ket{s}$ with detuning $\Delta$ and $\delta$, respectively.}
    \label{fig:3-level}
\end{figure}
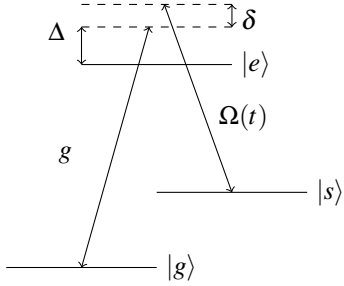

A quantum memory can be implemented by coupling two fields with a $\Lambda$-type three-level atom (or in fact any system that exhibits such level structure), comprising of a ground state $\ket{g}$, an excited state $\ket{e}$, and a metastable state $\ket{s}$ (see Fig.~\ref{fig:3-level}). The excited state is coupled to each of the two lower states via an electric field. To be more precise, the transition between $\ket{g}$ and $\ket{e}$ is coupled by a signal field with the coupling constant $g$ while the transition between $\ket{s}$ and $\ket{e}$ is coupled by a control electric field with the Rabi frequency $\Omega$.\\
\indent As this structure of a quantum memory is in a cavity of volume $V$, the composite state $\ket{\Psi}$ of atomic and cavity fields is written together in the dressed state,
\begin{equation}
    \ket{\Psi} = \ket{\text{atomic state}} \otimes \ket{\text{photon number}}.
\end{equation}
The Hamiltonian describing this composite system in the rotating frame is given by \cite{Gorshkov2007-1, Vitanov2017}
\begin{equation}
    H = \hbar\pqty{\Omega\hat{\sigma}_\text{se}\otimes\hat{1} + g\hat{\sigma}_\text{ge}\otimes \hat{a} +\text{H.c.}} + \Delta\hat{\sigma}_\text{ee}\otimes\hat{1} + \delta\hat{\sigma}_\text{ss}\otimes\hat{1},
    \label{eq:Hamiltonian}
\end{equation}
where the atomic operator $\hat{\sigma}_{ij}=\ketbra{i}{j}$; $\Delta$ is the one-photon detuning from the $\ket{g}-\ket{e}$ transition; and $\delta$ is two-photon detuning from the $\ket{g}-\ket{s}$ transition, which ideally should be negligible.\\
\indent The adiabatic quantum memory maps the initial state of the atom and one photon $\ket{g}\otimes\pqty{\alpha_0\ket{0}+\alpha_1\ket{1}}$ to the state of $\pqty{\alpha_0\ket{g} + \alpha_1\ket{s}}\otimes\ket{0}$. The transfer process is done via the \emph{stimulated Raman adiabatic passage} (STIRAP) process, which transfers along with the change of the dark state $\Phi_0$ \cite{Vitanov2017, Koerber2017}. The dark state $\ket{\Phi_0}$ is an instantaneous eigenstate with eigenvalue equal to zero,
\begin{align}
    \ket{\Phi_0} &= \cos \theta \ket{g,1} + \sin \theta \ket{s,0}\label{eq:dark-state-angle}\\
    \theta\ &= \arctan \frac{g}{\Omega(t)}.
\end{align}
\indent The mixing angle $\theta$ has to change sufficiently slowly for the wavefunction of this composite system to adjust itself and stay in the instantaneous eigenstate, i.e., the dark state. We assumed a control pulse $\Omega(t)$ to be a sigmoid curve with the time scale $T$ as stated in Eq.~\eqref{eq:control-pulse}. The change in $\Omega(t)$ can be controlled by the external electric field according to
\begin{align}
    \Omega (t) &= \bra{e}\hat{d}\cdot \epsilon_2 \ket{s} \frac{\mathcal{E}_2 (t)}{2\hbar},
\end{align}
while the coupling constant $g$ is kept constant according to 
\begin{equation}
    g = \bra{e}\hat{d}\cdot \epsilon_1 \ket{g} \sqrt{\frac{\omega_1}{2\hbar \epsilon_0 V}},
    \label{eq:couplingConstant-g}
\end{equation}
where $V$ is a volume of the cavity; $\hat{d}_{ge}$ is a transition dipole moment between pair of states $\ket{g}$ and $\ket{e}$; $\mathcal{E}_2 (t)$ is an envelope of the control field; and $\epsilon_1$ and $\epsilon_2$ are polarization unit vectors of the signal and control fields, respectively \cite{Gorshkov2007-1}. Note that this work also showcases modeling in hBN defects. In principle, in the calculation of the transition dipole moment, the refractive index of hBN needs to be included, which would result in an integral of the moment over the hBN thickness. However, such thickness is comparatively small compared to the optical wavelength. As such, the effects from the refractive index of hBN (1.85 according to Ref.~\cite{10.1021/acsphotonics.8b00127} in the visible spectrum) are then negligible.\\
\indent From the Hamiltonian in Eq.~\eqref{eq:Hamiltonian}, the dynamics of a quantum memory are simulated numerically in a scaled unit with coupling constant $g$.  Once the final state of the atom is obtained, we can calculate how the memory performs. The decay mechanism is phenomenologically modeled by the apparent population decay rates and dephasing rates. The Lindblad master equation which takes these decay rates into account is given by
\begin{equation}
    \frac{d}{dt}\rho = -i[H,\rho] + \sum_{i,j} C_{ij}\rho C^\dagger_{ij} - \frac{1}{2} \anticommutator{C^\dagger_{ij} C_{ij}}{\rho},
    \label{eq:masterEq_decay}
\end{equation}
where $C_{ij} = \sqrt{\gamma_{ij}}\ketbra{i}{j}$ denotes a jump operator. Whenever $i=j$, the decay rates $\gamma_{ij}$ describe a purely dephasing process. Note that the decay rate $\gamma_{ij}$ can be obtained from experiments or first-principle calculations, such as density functional theory (DFT), by considering the spin-allowed transition energy between two states $\ket{i}$ and $\ket{j}$ under the Fermi's golden rule as a constraint \cite{C9TC02214G}. 

\subsection{Performance metrics}
The performance of a quantum memory is evaluated via  four criteria, namely, fidelity, efficiency, storage time and bandwidth \cite{Lvovsky2009, Simon_2010, Heshami2016}. 

In essence, the quantum state fidelity measures how close the quantum state is before storage and after retrieval. However, we did not directly calculate it in this work as we have not yet embedded any quantum information into the qubit. Moreover, all quantum states in the numerical calculations without the decay rates are always pure. This reduces fidelity to coincide with efficiency.\\
\indent To that end, the efficiency can be thought of as the probability of successfully performing the task, and we can distinguish it into three types; namely, writing efficiency, reading efficiency, and global (overall) efficiency. The writing efficiency is the probability of successfully mapping the photonic state to the atomic state; vice versa, the reading or retrieving efficiency is that from mapping the atomic state to the photonic state; and the global efficiency is the probability of successfully carrying out the whole process, which also includes the effects of storage in a cavity. To be precise, the writing efficiency $\eta_w$ is
\begin{equation}
\eta_w = \lim_{t\to \infty} \bra{s}\rho(t)\ket{s},
\end{equation}
and analogously the reading efficiency is
\begin{equation}
\eta_r = \lim_{t\to \infty} \bra{g}\rho(t)\ket{g}.
\end{equation}
Both processes have their initial condition $\rho(t=0)$ be the empty and stored states, respectively, i.e. $\rho(t=0)=\ketbra{g}{g}$ for the writing process and $\rho(t=0) = \ketbra{s}{s}$ for the reading process.
The limit $t \rightarrow \infty$ indicates that the transfer of a quantum state from the initial state to the final state adiabatically takes place. However, only processes with finite time are physically feasible. Thus, we examined over a certain time in which the adiabatic path of the process yields 99\% of the asymptotic population in the state. For this purpose, the population from the adiabatic path is derived from Eq. \eqref{eq:dark-state-angle}, and can be expressed as
\begin{equation}
    \textrm{population of } \ket{g,1} = \cos ^2 \theta = \frac{\Omega^2 (t)}{\Omega^2 (t) + g^2 }.
\end{equation}
We emphasize that the writing or reading efficiency calculated in this work does not include the storage efficiency; it is merely the transition probability from the $|s\rangle$ state to the $|g\rangle$ state. As such, we did not calculate the overall efficiency as it would require the consideration of storage time. Moreover, we assume that the reading process is a time-reversed process of the writing process.

\indent Other than the controllable parameters (parameters related to controlling the memory), the decay rates also play roles in both the mapping (writing and reading) and global efficiencies of the quantum memory. Specifically, we show in Section \ref{sec:decay-dephasing-eff} that the decay rates from states $\ket{s}$ to state $\ket{g}$, pure dephasing rate of $\ket{g}$, and pure dephasing rate of $\ket{s}$ affect the mapping efficiencies.\\
\indent The storage time is defined as how long a quantum state can be stored. We remark that while the storage time of a quantum memory is a primary factor for determining the quality of a quantum memory, and is usually used for comparison between performances of different quantum memory units, we believe that other performance metrics as mentioned above should be taken into account as well. In addition, the bandwidth yielding the range of frequencies that can be used in such a quantum memory should be also considered \cite{Simon_2010, Heshami2016}, as it determines the integrability of the quantum memory with other components in a quantum network.\\
\indent We additionally remark that four-wave mixing noise is ignored in our modelling. Even though the interaction of the control field can introduce this noise, it largely depends on the electronic structures and polarization of the transition among particular defect states. Thus, it strongly depends on each defect type, which is out of the scope of this work. When a particular defect is evaluated with a specific control field, it should be more beneficial to also evaluate the effects of four-wave mixing noise.

\section{Results and discussion}
\subsection{Performance metrics}
\label{Performance-scale}
\subsubsection{Efficiencies in the absence of decay}
\begin{figure}
    \centering
        \centering
        \includegraphics[width = .5\textwidth]{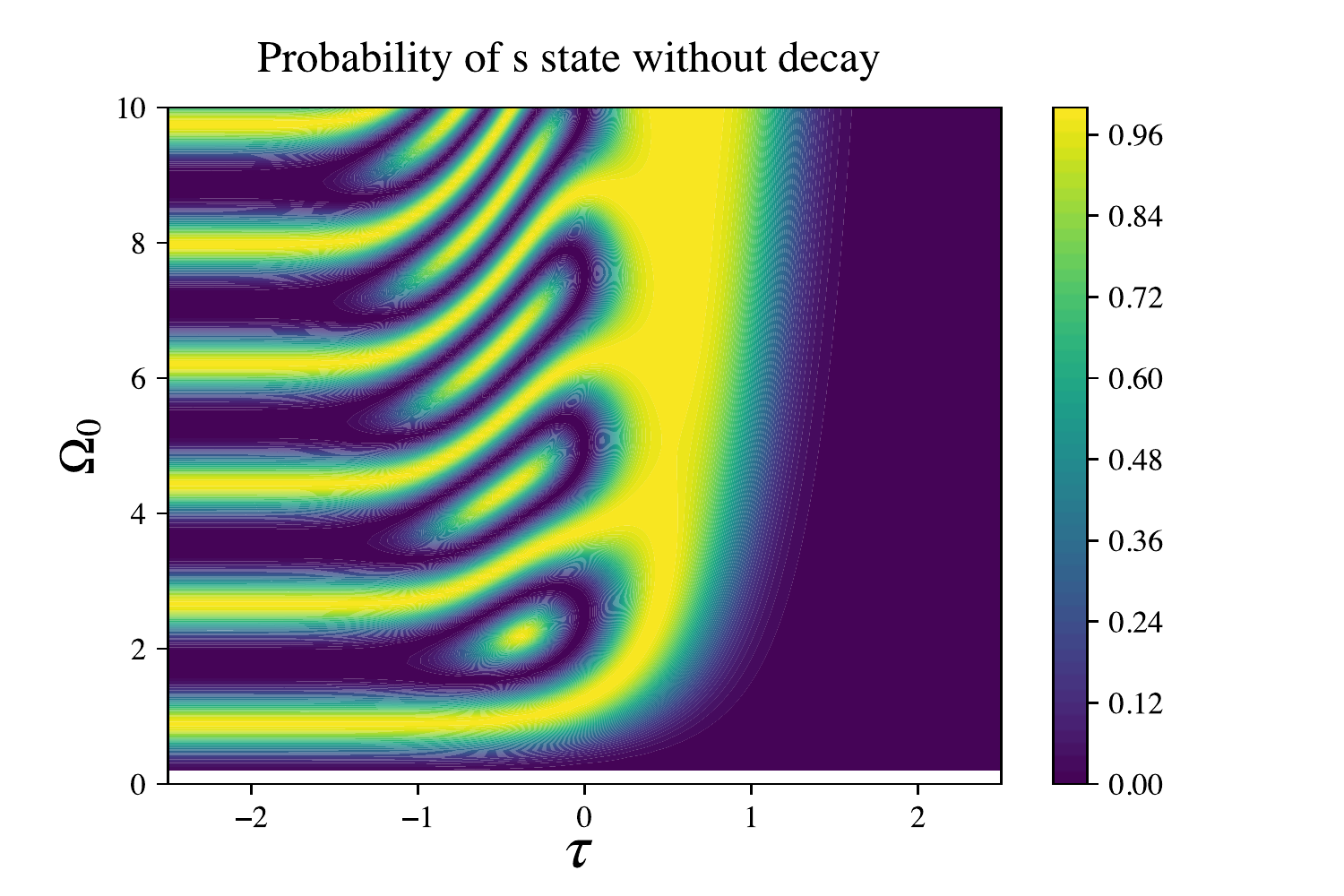}
        \caption{Transition probability to the metastable state $\ket{s}$ for Gaussian pulses of $g(t)$ and $\Omega(t)$ in each pair of control parameters benchmarked with Ref. \cite{Vitanov2017}.}
        \label{fig:probS-Gaussian}
\end{figure}

\begin{figure}
    \centering
    \begin{subfigure}[b]{0.49\textwidth}
        \includegraphics[width=\textwidth]{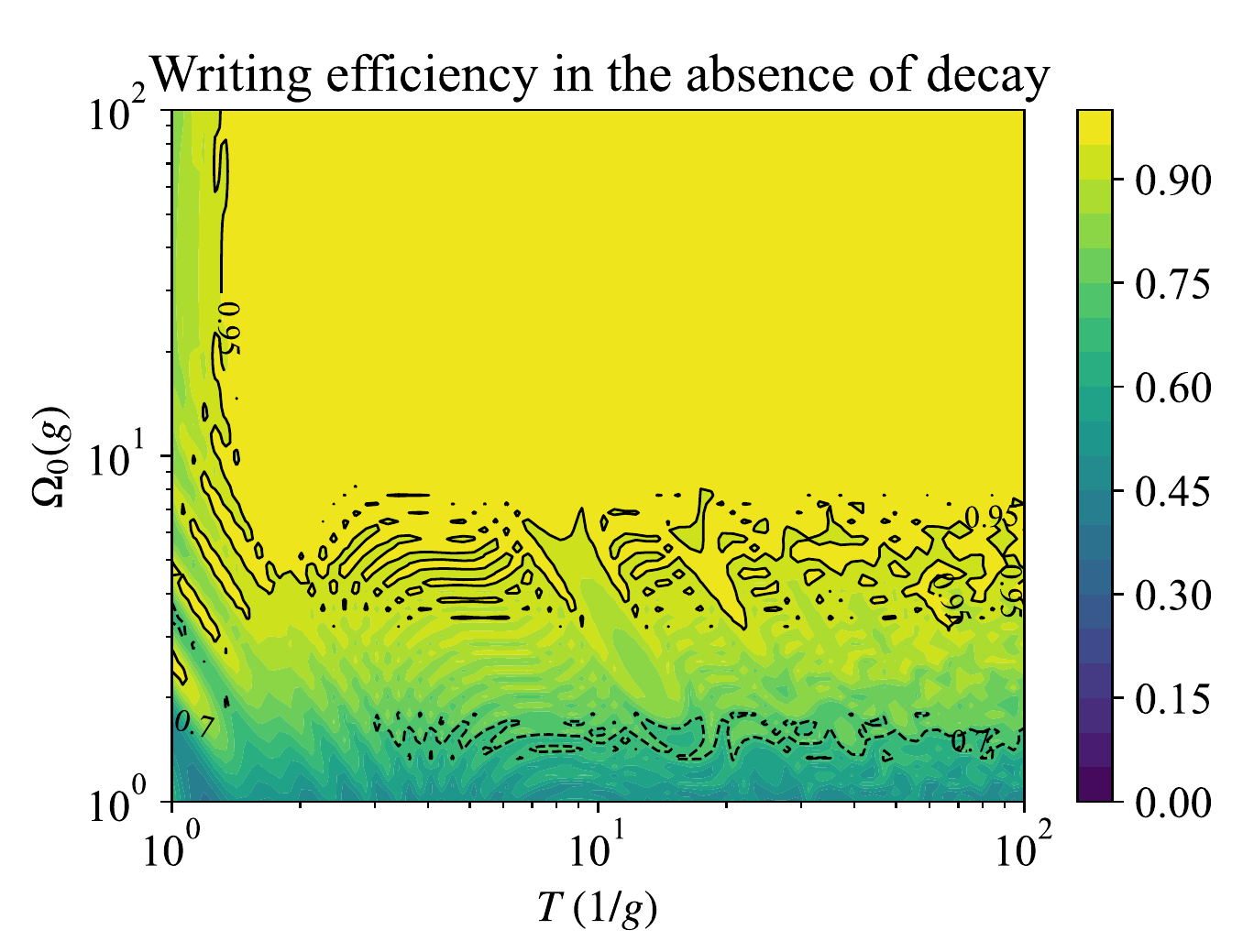}
    \end{subfigure}
    \begin{subfigure}[b]{0.49\textwidth}
        \includegraphics[width=\textwidth]{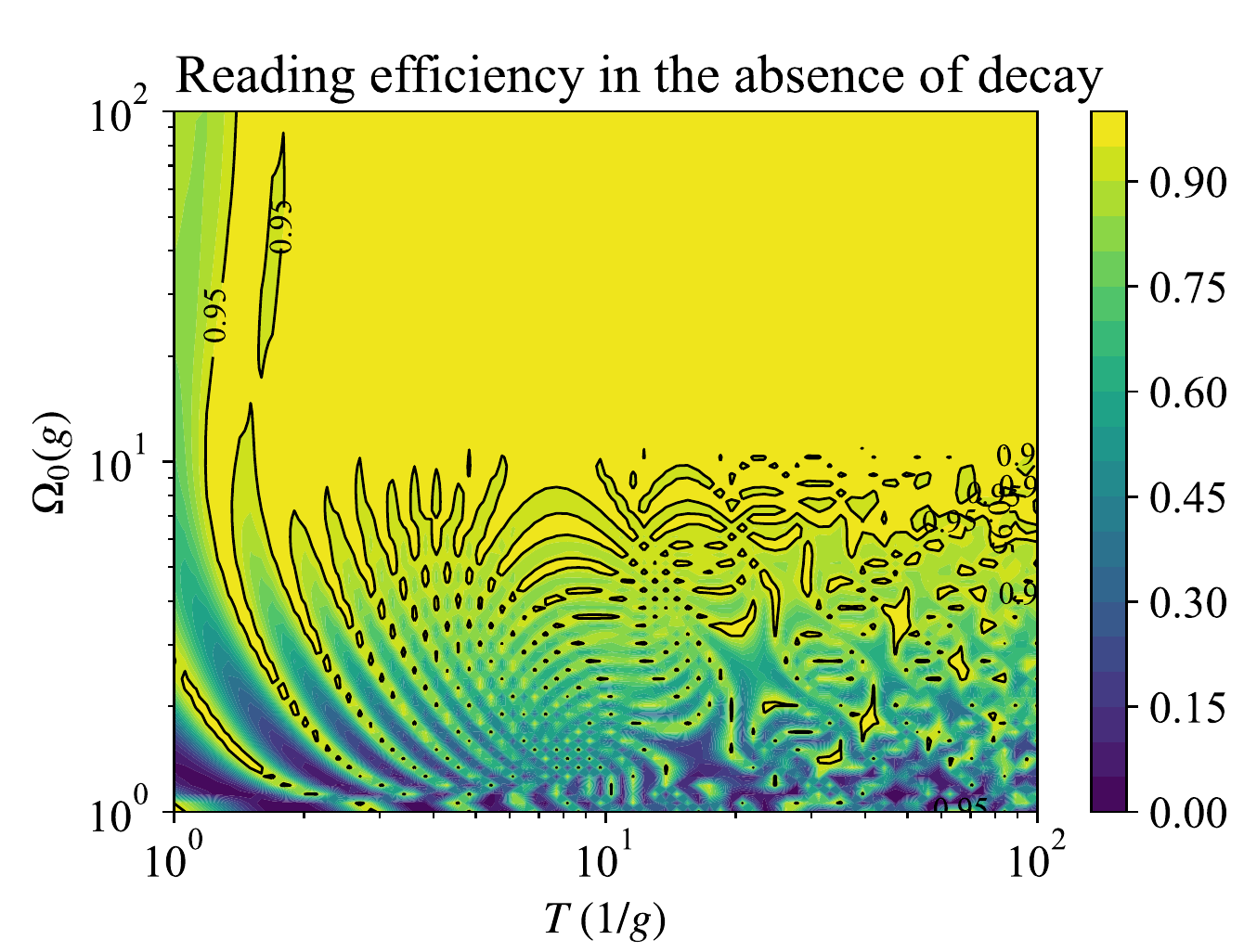}
    \end{subfigure}
    \caption{Writing and reading efficiencies for different $\Omega_0$ and $T$ in log-log scale. The slower characteristic time and higher control field strength yield high writing and reading efficiencies. The efficiencies reach $\geq 95\%$ as the field strength $\Omega_0$ is larger than $10g$.}
    \label{fig:contour-efficiency-writing}
\end{figure}

For the writing (resp.\ reading) process, we simulated the dynamics from the Hamiltonian $H$ in Eq.~\eqref{eq:Hamiltonian} to obtain the probability of transition from the initial state $\ket{g}$ (resp.\ $\ket{s}$) to the final state $\ket{s}$ (resp.\ $\ket{g}$). The simulation is implemented using Python code and the QuTiP package \cite{Johansson2013}.\\
\indent To verify our numerical methodology, we first benchmarked the simulation of semi-classical STIRAP with work by Vitnanov \emph{et al.} \cite{Vitanov2017}. The semi-classical STIRAP assigns another controllable classical electric field instead of the signal field. Both control fields $g(t)$ and $\Omega(t)$ have pulses in the Gaussian shape as given below
\begin{align}
    \Omega(t) &= \Omega _0 e^{-(t+\tau/2)^2} \\
    g(t) &= \Omega _0 e^{-(t-\tau/2)^2}.
\end{align}
\indent Our result as shown in Fig.~\ref{fig:probS-Gaussian} agrees well with their work although the ripple pattern in our vertical axis is twice as dense as that of them. This can be accounted for by the half of $\Omega$ magnitude defined.\\
\indent Considering the writing/reading efficiency landscape as a function of $\Omega_0$ and $T$, Fig.~\ref{fig:contour-efficiency-writing} indicates that the slower mapping time and larger control strength result in a better performance of the writing/reading process. The figure also suggests that the $95\%$ writing/reading efficiency can be reached by the conditions $T > 2/g$ and $\Omega_0 > 10$. In the following section, we only consider the writing process.\\
\indent In a more realistic model considering the decay rates in both atomic and cavity fields, the global efficiency is then not directly a product of only both efficiencies because it has to take the memory time into account as the population stored in $\ket{s}$ changes. Also, the characteristic time $T$ cannot be extended indefinitely due to the decay and dephasing rates of the material and the cavity field. Moreover, if the transition energy of a photon source and the photon frequency do not match the memory requirements, then the detuning is non-zero, hence reducing efficiency. We shall discuss these issues in the following section.\\

\subsubsection{Electronic decay rates and efficiencies}
\begin{figure}
    \center
    \begin{subfigure}[b]{0.47\textwidth}
        \includegraphics[width=\textwidth]{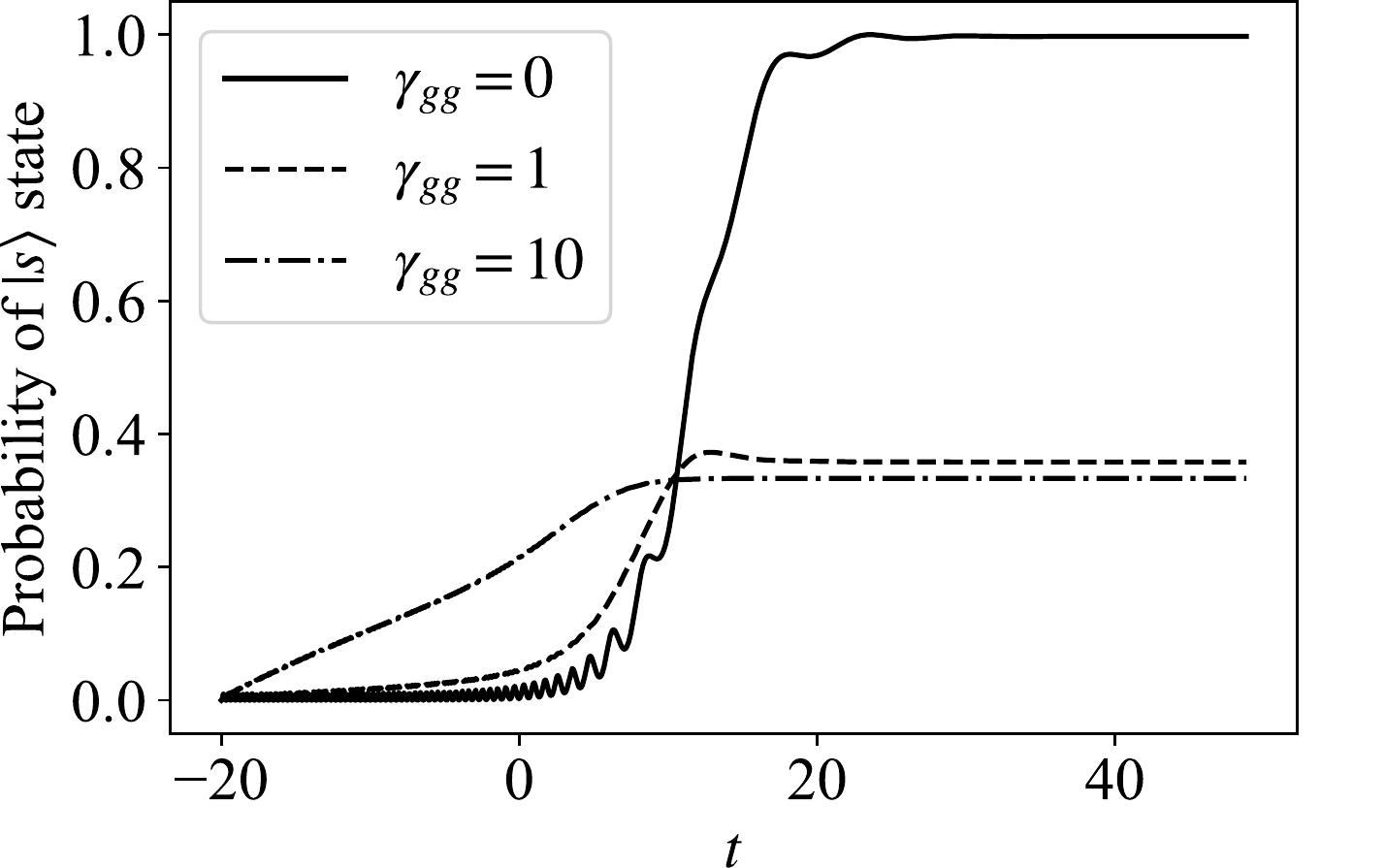} 
        \label{fig:dephasing-dynamics}
    \end{subfigure}
    \begin{subfigure}[b]{0.46\textwidth}
         \includegraphics[width=\textwidth]{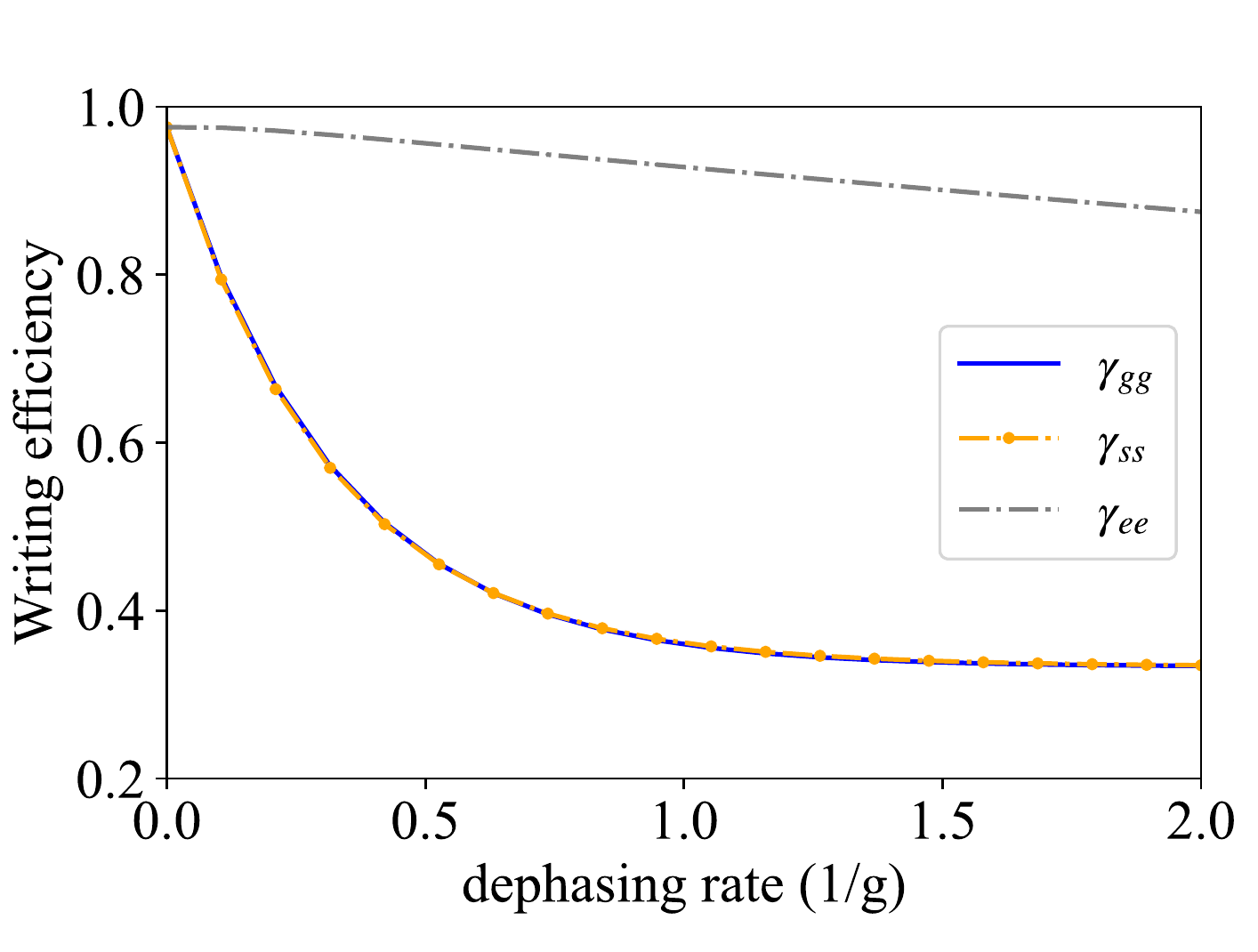}
        \label{fig:dephasing_gse}
    \end{subfigure}
    \begin{subfigure}[b]{0.45\textwidth}
        \includegraphics[width=\textwidth]{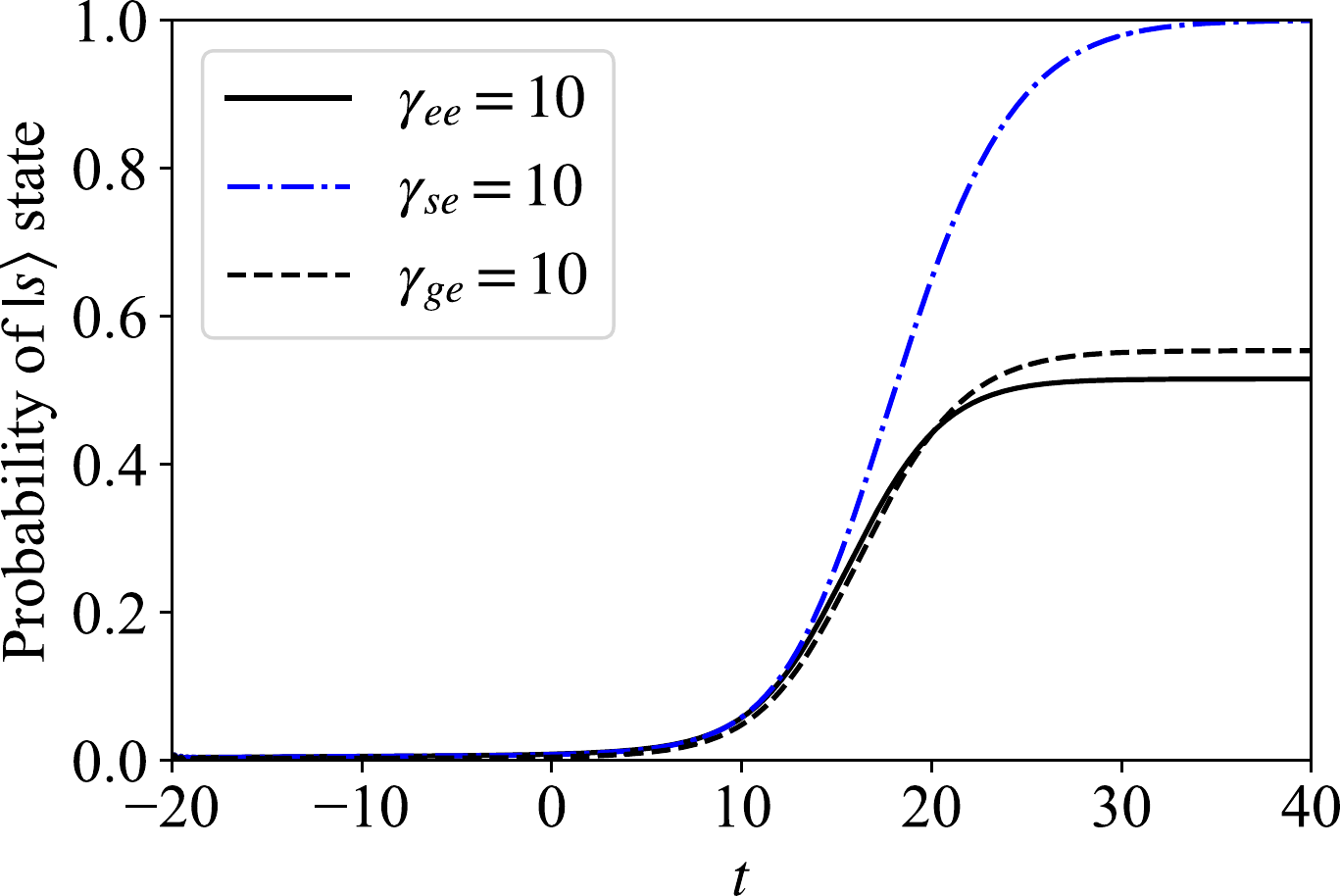}
        \label{fig:dephasing_g-eff}
    \end{subfigure}
    \caption{Dynamics with dephasing over dimensionless time $t$. (top) Probability of transferring a quantum state to the $|s\rangle$ state as a function of time $t$. (middle) Writing efficiency as a function of dephasing rates. Dephasing in each of three states and a resulting probability of state $|s\rangle$.  (bottom) Probability of transferring a quantum state to $|s\rangle$ state with different decay rates relevant to the excited state. The decay rates in the bottom figure are in the relative unit $g$.}
    \label{fig:dephasing_dynamics_and_Prob_s}
\end{figure}

To investigate if the decay rates in our model can affect the adiabaticity, we have to transform the decay rate from the basis of $\ket{g}$, $\ket{s}$, and $\ket{e}$ to that of $\ket{\Phi_0}$ (dark state), $\ket{\Phi_-}$ and $\ket{\Phi_+}$ (bright states). The transformation matrix is constructed from Eqs.~\eqref{eq:phi0}, \eqref{eq:phi_p-non_zero_Delta}, and \eqref{eq:phi_m-non_zero_Delta} stated in the Appendix~\ref{sec:adiabatic-appendix}. For example, to transform the decay rate from $\ket{g}$, the state of $\ket{g}$ is projected into the basis of $\{\ket{\Phi_0},\ket{\Phi_-}, \ket{\Phi_+}\}$ as
\label{sec:decay-dephasing-eff}
\begin{align}
     \ket{g} &= \braket{\Phi_0}{g}\ket{\Phi_0} + \braket{\Phi_-}{g}\ket{\Phi_-} + \braket{\Phi_+}{g}\ket{\Phi_+} \\
     &= \cos\theta\ket{\Phi_0} + \sin\theta\cos\phi\ket{\Phi_-} + \sin\theta\sin\phi\ket{\Phi_+}.
\end{align}
Then, the projection $\ketbra{g}{g}$ can be expressed in the adiabatic basis as shown in Appendix~\ref{sec:Projection-in-adiabatic}. This also includes all projections of $\ket{g},\ket{s},$ and $\ket{e}$.\\
\indent These expressions of the jump operators on an adiabatic basis can inform us which decay rate is relevant in determining the efficiency of the quantum memory. We note that the jump operators here are written in the basis order of $\ket{\Phi_0}$, $\ket{\Phi_-}$, and $\ket{\Phi_+}$. The non-adiabatic contribution comes from off-diagonal elements of the first row and column. This suggests that the relevant decay rates are all electronic decay rates except the dephasing rate of the excited state. Note that we did not consider the \emph{spontaneous excitation} as it would be relatively small in realistic systems.\\
\indent We then verified our findings from analytic expressions with numerical calculation using QuTiP package \cite{Johansson2013}. Figs.~\ref{fig:dephasing_dynamics_and_Prob_s} (top and middle) show that the dephasing rate in $\ket{g}$ affects the writing efficiency (mapping from $\ket{g}\rightarrow\ket{s}$), and the probability of successful transfer, but this probability saturates after time $t>20$, which can be seen in Fig.\ref{fig:dephasing_dynamics_and_Prob_s} (top). Fig.~\ref{fig:dephasing_dynamics_and_Prob_s} (middle) shows that the larger dephasing rate of $\gamma_{gg}, \gamma_{ss},$ or $\gamma_{ee}$ results in lower writing efficiency. In addition, it is evident from Fig.~\ref{fig:dephasing_dynamics_and_Prob_s} (bottom) that all the dephasing rates are relevant, including $\gamma_{ee}$. However, the $\gamma_{ee}$ dephasing rate does not affect the writing efficiency as strongly as other dephasing rates; see Fig.~\ref{fig:dephasing_dynamics_and_Prob_s} (middle).\\

\subsubsection{Cavity field decay}
\label{sec:cavityFieldDecay}
\begin{figure}
\centering
\begin{subfigure}[b]{0.5\textwidth}
        \includegraphics[width = \textwidth]{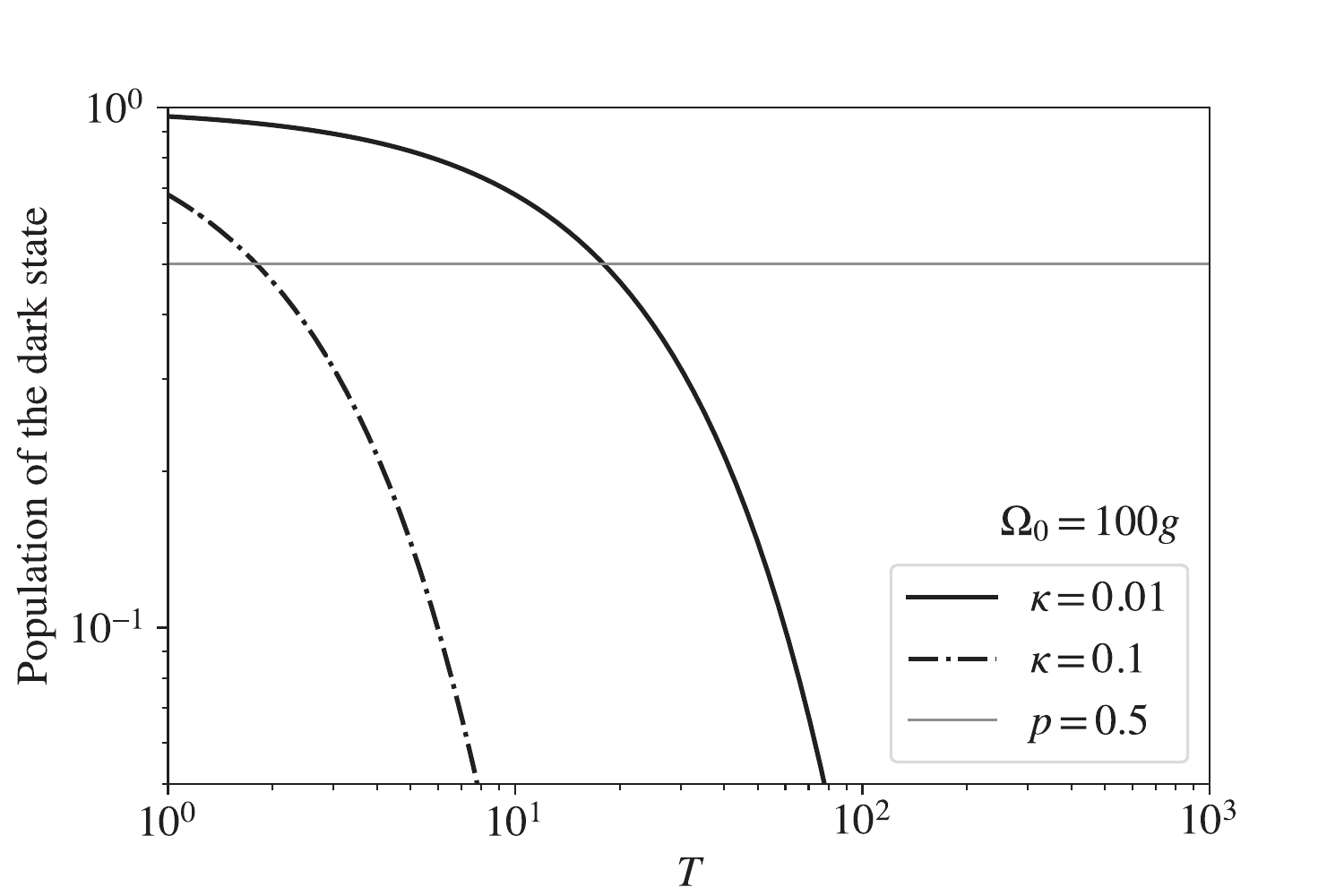}
\end{subfigure}
\caption{Population of the dark state in the presence of different cavity field decay rates $\kappa$, where the line $p=0.5$ indicates the regime in which a quantum memory can in principle beat the no-cloning limit when the system efficiency is above 50\%. $\kappa$ and $T$ are in the unit system in which $g=1$.}
\label{fig:cavity-field-decay-population}
\end{figure}

To consider the limit of how much we can set the characteristic time $T$, the decay rate of a field in a cavity $\kappa$ has to be taken into account. This is related to the quality factor $Q$ of the cavity by
\begin{equation}
    \kappa = \frac{\omega}{2 Q},
    \label{eq:decay-quality}
\end{equation}
where 
\begin{equation}
    Q \equiv \frac{\omega_0}{\Delta\omega}.
\end{equation}
Here $\omega_0$ is the resonance frequency of the cavity, and $\Delta\omega$ is its full width at half maximum (FWHM) of the cavity response curve. The non-zero value of $\kappa$ results in a loss of photons in the cavity, and consequently limits how long the mapping process can take. The loss is described by the process of mapping the photon number states of $\ketbra{1}{1}$ to $\ketbra{0}{0}$:
\begin{equation}
\ketbra{1}{1}\rightarrow e^{-\kappa t} \ketbra{1}{1} + (1-e^{-\kappa t}) \ketbra{0}{0}.
\end{equation}
If the time scale parameter $T$ is of comparable size to $1/\kappa$, then more photons leak out of the cavity before the writing process ends.\\
\indent To investigate quantitatively how $\kappa$ limits the characteristic time $T$, we considered a perfectly adiabatic quantum memory operation with zero decay rate into other modes. That is, the dissipation term in Eq.~\eqref{eq:masterEq_decay} only has one $C_{ij}$ describing the decay of the cavity field which $\ket{i} = \ket{g,0}$ and $\ket{j} = \ket{g,1}$. This state $\ket{g,1}$ is what we have written in a reduced form into $\ket{g}$, while the state $\ket{g,0}$ is neglected in the previous analysis. We shorten this vacuum state to $\ket{v}$, representing a ground state dressed with the \emph{vacuum} state. When the adiabatic condition is assumed, the Hamiltonian and density operators are always diagonal. Details of the off-diagonal elements can be found in the Appendix~\ref{sec:adiabatic-appendix}. Then, the equation of motion can be reduced to contain only two terms of the dark state population $d(t)$ and the ground-vacuum state, $\ket{g,0}$, population $v(t)$. First, the decay rate in the dark state due to the cavity field decay $\kappa_a$ is 
\begin{equation}
    \kappa_a(t) = \kappa \cos^2\theta(t).
\end{equation}
The system of equations governing this decay in the adiabatic process is 
\begin{align}
    \dv{}{t} d(t) &= -\kappa_a(t) d(t) = -\kappa \cos^2\theta(t) d(t) \label{eq:ode1},\\
    \dv{}{t} v(t) &= \kappa_a(t) d(t) \label{eq:ode2}.
\end{align}
Assuming the control field $\Omega (t)$ is given by Eq.~\eqref{eq:control-pulse}, then Eq.~\eqref{eq:ode1} can be solved analytically for a general solution and it is given by
\begin{eqnarray}
        d(t) =c_1 &\exp& \pqty{\frac{\kappa\Omega _0^2 \left(T \log \left(\left(e^{t/T}+1\right)^2+\Omega _0^2\right)-2 t\right)
        }{2 \left(\Omega _0^2+1\right)}} \nonumber \\
        \times &\exp& \left(\frac{2\kappa  \Omega _0 T \tan ^{-1}\left(\frac{e^{t/T}+1}{\Omega _0}\right)}{2 \left(\Omega _0^2+1\right)}\right)
\end{eqnarray}

\noindent where $c_1$ is a constant, depending on the initial condition, which is set as $d(t_0)=d_0$ at a particular $t_0$. Suppose that $d_0 = 0.999$, then the initial time $t_0$ is approximately when to put a photon into a cavity. Here, we assume that a photon enters the cavity when $99.9\%$ of the dark state population is in $\ket{g}$. To keep time $t$ in Eq.~\eqref{eq:control-pulse} the same, the initial time $t_0$ is set according to
\begin{equation}
    t_0 = T \log \pqty{\frac{p \abs{\Omega_0} }{\sqrt{p(1-p)}}-1},
    \label{eq:time_in_sigmoid}
\end{equation}
where $p$ is the probability of the dark state overlapping a metastable state.\\
\indent The value $t$ for $p = 0.001$ can be taken as an initial time $t_0$. From this, we calculated values of the constant $c_1$ and solved for $d(t)$ at subsequent time and examined its behavior as $t\rightarrow \infty$. The plot of the population versus the characteristic time $T$ is shown in Fig.~\ref{fig:cavity-field-decay-population}, in which we found that as $T$ increases, the writing efficiency is reduced. We annotate a horizontal line of efficiency $p=0.5$ to indicate the lower bound for a quantum memory in principle to have a chance to beat the no-cloning limit (overall efficiency has to be greater than 50\%), which the memory has to pass to be useful in quantum networks \cite{Cho2016,Grosshans2001}. For the cavity decay rate $\kappa$ in the order of $0.01$, the maximum $T$ which yields a final population of the dark state $p>0.5$ is in the order of 10. Similarly, if the decay rate $\kappa$ is in the order of $0.1$, the maximum $T$ with $p>0.5$ will be in the order of $1$. As indicated in Fig.~\ref{fig:contour-efficiency-writing} that the regime where $T>2$ has the optimal efficiency, we can conclude that the high cavity decay rate may not reach the optimal efficiency. Hence, the cavity field decay rate restricts the range of $T$ and how efficient the memory is. \\
\indent As a backbone of security in quantum communication, the no-cloning limit is pertaining to the distinguishability between a real photon from a quantum memory and that produced by an attacker provides advantages over classical communication \cite{Cho:16}. The chance of an eavesdropper intercepting and reproducing a state such that the sender (Alice) and receiver (Bob) cannot detect the tampering is $50\%$. If the quantum memory fails to emit back the photon at a rate greater than $50\%$ of the time, Alice and Bob cannot tell if there is someone else tampering with the quantum channel. As such, the overall efficiency must be $>50\%$ to be useful in a quantum network, which requires the writing efficiency to be $>50\%$. We note that we can only show the writing efficiency since the overall efficiency has to take storage time into account, which is beyond the scope of our work. 

\subsubsection{Bandwidth}
\begin{figure}
    \begin{subfigure}[b]{.49\textwidth}
        \includegraphics[width=\textwidth]{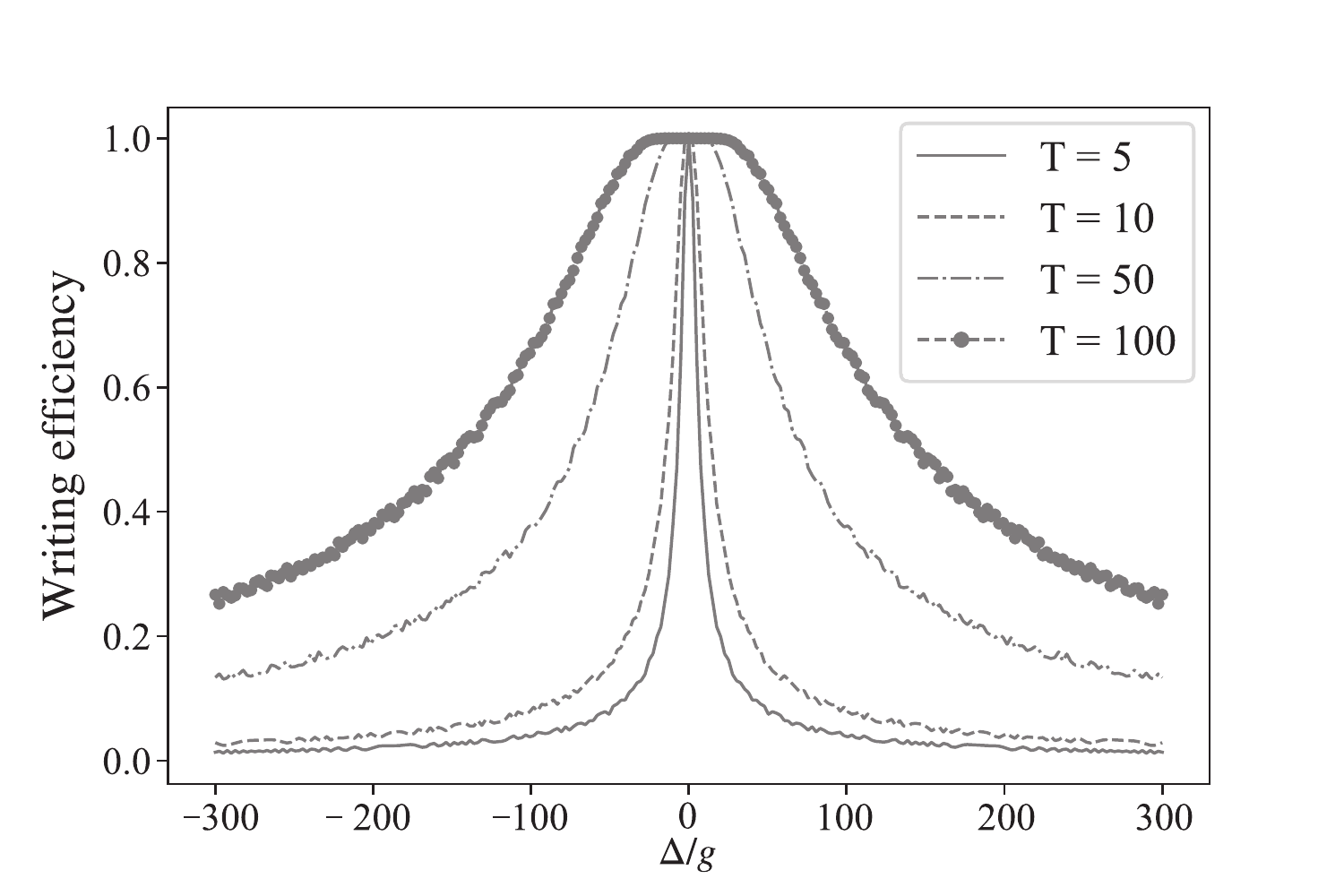}
        \caption{}
        \label{fig:2-photon-absorption-spectrum-detuning-100}
    \end{subfigure}
    \hfill
    \begin{subfigure}[b]{.49\textwidth}
        \includegraphics[width=\textwidth]{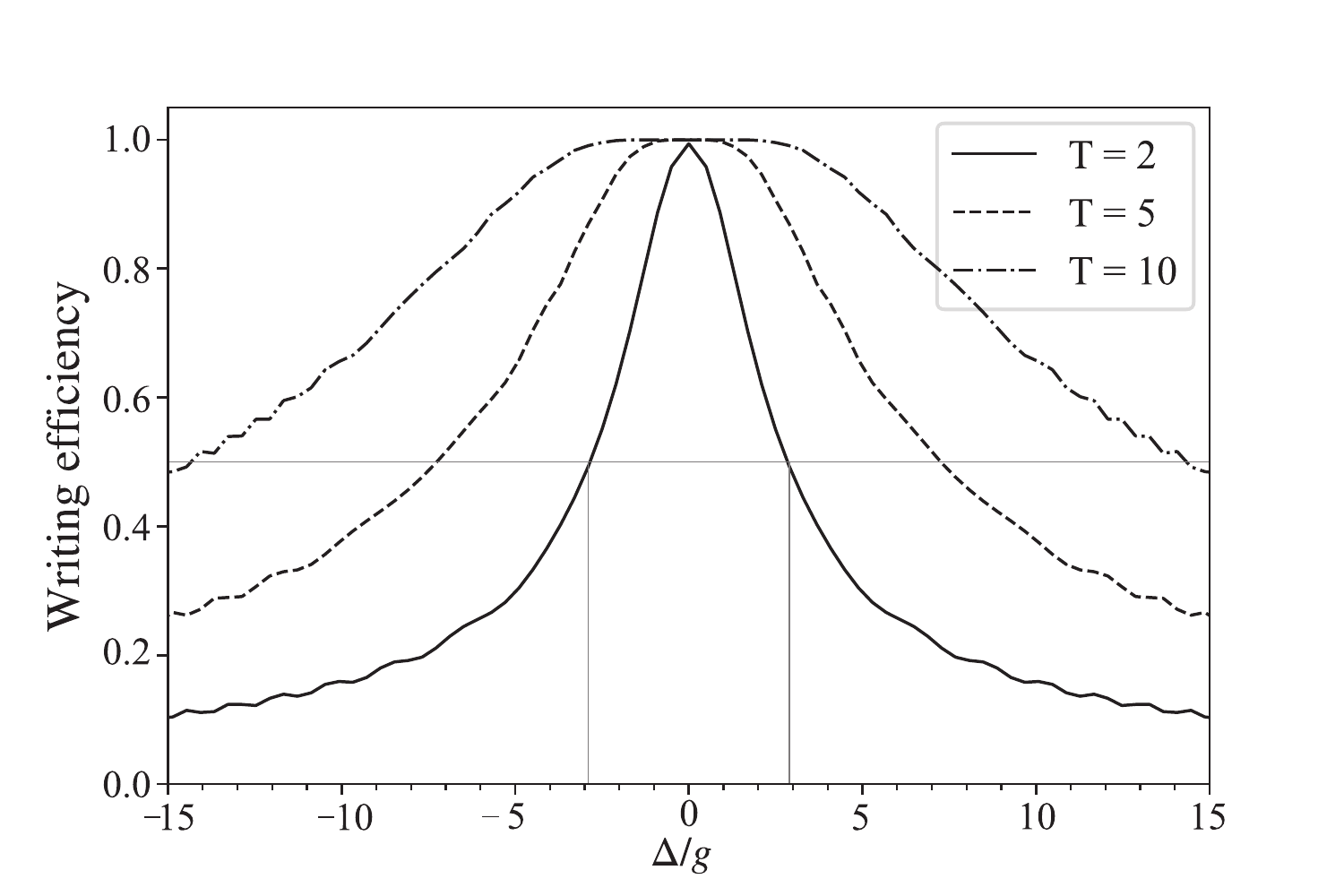}
        \caption{}
        \label{fig:2-photon-absorption-spectrum-detuning-15}
    \end{subfigure}
    \caption{Writing efficiency and detuning resonance for different characteristic time $T$ in the unit of $g=1$ where (a) $\Omega_0=100$ while (b) is the zoomed-in perspective.}
    \label{fig:2-photon-abs-spectrum}
\end{figure}
It is expected to see a reduction of mapping efficiencies once signal and control field have non-zero detuning with their corresponding transition. Although the desired transitions are between $\ket{g}$ and $\ket{s}$, the detuning between $\ket{g}$ and $\ket{e}$ reduced the energy gaps between instantaneous eigenstates (see Sec.~\ref{sec:non-adiabatic-correction}) and the condition of slow change in $\theta$ is even slower.\\
\indent Fig.~\ref{fig:2-photon-absorption-spectrum-detuning-100} shows the writing efficiency with one-photon detuning $\Delta$ when using Gaussian pulses where the curves show a finite width absorption window around the resonance. This pinpoints that the longer characteristic time $T$ allows a wider bandwidth. Their relation can be mathematically proven as follows.\\
\indent As $g \propto \sqrt{\omega_1}$ as stated in Eq.~\eqref{eq:couplingConstant-g}, then $g$ and $\Delta$ depend on one another as
\begin{equation}
    g\propto \sqrt{\omega_1} \longrightarrow g = C\sqrt{\omega_1} = C \sqrt{\omega_\text{ge} + \Delta}.
    \label{eq:couplingConstant-g_omega1_relation}
\end{equation}
\indent From the estimate of half width at half maximum $\sigma_\Delta$ in a physical unit, it follows that 
\begin{equation}
    \sigma_\Delta \equiv \text{HWHM} \equiv\frac{\Delta }{g} = \frac{\Delta}{C \sqrt{\omega_\text{ge} + \Delta}},
\end{equation} 
where $\sigma_\Delta$ is a half-width half maximum obtained from numerical calculations. By choosing a transition frequency $\omega_\text{ge}$, and solving a simple quadratic equation, one can find $\Delta$ at half maximum. This equation has a general solution in the form
\begin{equation}
    \Delta = \frac{1}{2} \pqty{C^2 \sigma_{\Delta} ^2 \pm \sqrt{C^4 \sigma_{\Delta} ^4+4 C^2 \sigma_{\Delta}^2 \omega_\text{ge}}},
    \label{eq:Bandwidth_quadratic_formula}
\end{equation}
where by Eq.~\eqref{eq:couplingConstant-g},
\begin{equation}
    C = \frac{\bra{e}\hat{d}\cdot\vec{\epsilon}\ket{g}}{\sqrt{2\hbar\varepsilon_0 V}} = \frac{g}{\sqrt{\omega_1}}.
\end{equation}
\begin{figure*}[ht]
    \includegraphics[width=1\textwidth]{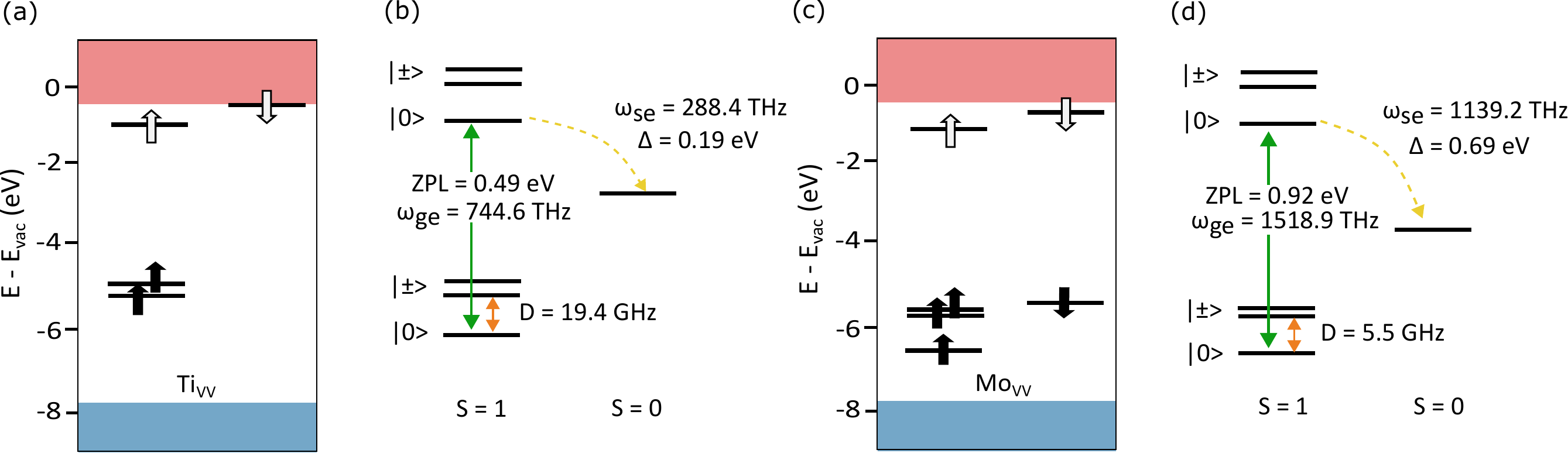}
    \caption{Electronic transition of Ti$_\text{VV}$ and Mo$_\text{VV}$ adapted from Ref. \cite{Smart2021} calculated using G$_{0}$W$_{0}$ approach with PBE0($\alpha$) initial wavefunctions. (a) and (c) are single-particle electronic structures of triplet configuration where the blue and red regions illustrate valence and conduction bands, respectively. The filled and unfilled arrows represent the occupied and unoccupied defect states, indicating spin direction by up and down. (b) and (d) depict the associated electronic transition of Ti$_\text{VV}$ and Mo$_\text{VV}$ with an intersystem-crossing channel between triplet (S=1) and singlet (S=0) configurations. The frequencies $\omega_{ge}$ and $\omega_{se}$ are computed by this work.}
    \label{fig:Ti-VV_in_hBN}
\end{figure*}

\indent The conjugate root with a minus sign is neglected because $C^2\sigma_{\Delta}^2 < \sqrt{C^4\sigma_{\Delta}^4+4C^2\sigma_{\Delta}^2\omega_\text{ge}}$. A simple approximation shows that if the detuning $\Delta/g$ is small compared to $\omega_\text{ge}$, then Eq.~\eqref{eq:Bandwidth_quadratic_formula} is reduced to a simple relation
\begin{equation}
    \Delta = C \omega_\text{ge} \sigma_\Delta = g \sigma_\Delta.
    \label{eq:simplify-bandwidth}
\end{equation}

\indent In essence, the relation between bandwidth and the other parameters is approximately
\begin{equation}
    \Delta \propto \abs{\text{dipole transition matrix}}, \frac{1}{\sqrt{V}}, \sqrt{\omega_\text{ge}} \text{, and } \sigma_{\Delta}
\end{equation}
\indent From this relation, if the cavity has a small volume or long characteristic time $T$ while the material has a large dipole moment or high $\omega_{ge}$ (high ZPL), then the memory will have a high bandwidth for a range of frequencies.\\

\subsection{Application of hBN defects as quantum memory}
\label{example-defects}
To demonstrate a physical system feasible for being a quantum memory and evaluate its performance, we selected defects hosted by hBN as an example for the following reasons: (i) there are many defect choices emitting polarized light with the emission wavelengths covering from UV to NIR, allowing one to match a transition energy \cite{Cholsuk2022, anand-dipole}. (ii) 2D materials do not obstruct the collection of single photons \cite{Tran2016}. (iii) hBN material can operate under room temperature due to a large band gap of hBN. (iv) hBN is likely to have less electron-phonon coupling than other solid-state systems, leading to a high quantum efficiency. Lastly, (v) it has been reported earlier that some hBN defects inherit the consistent emission wavelength with quantum memory systems, resulting in a high coupling efficiency with other systems \cite{Cholsuk2022}. In this section, we aim to explore the correlation between control parameters (coupling constant, bandwidth, quality factor) and intrinsic defect properties.\\
\indent The three-level electronic states as depicted in Fig.~\ref{fig:3-level} stem from the defects localized in host hBN. Since those three-level states need to intrinsically behave as the desired $\Lambda$ structure, a defect needs to form two spin configurations, namely a singlet and a triplet. This will likely inherit the ground, excited, and meta-stable states as depicted in Fig.~\ref{fig:3-level}. For a triplet state, the transition between an occupied defect state ($\ket{g}$) and an unoccupied defect state ($\ket{e}$) should be radiative, so that the transition dipole moment is finite. The meta-stable state ($\ket{s}$) is generated by the singlet-spin state where the transition non-radiatively takes place between $\ket{e}$ and an unoccupied defect state ($\ket{s}$). It should be noted that other $\Lambda$ structure schemes are also possible \cite{Heshami2014}. Considering each transition, the transition energy is ideally expected to be in a similar range with another coupling component whilst the lifetime is desired to be sufficiently long for holding the photon states.\\
\indent Fig.~\ref{fig:Ti-VV_in_hBN} illustrates the electronic transition of two defect candidates in hBN, namely Ti$_{\text{VV}}$ and Mo$_{\text{VV}}$. These defects are bi-vacancies with an impurity. As revealed by Ref.~\cite{Smart2021}, both defects inherit triplet and singlet spin configurations, allowing for a possible three-level system via the intersystem-crossing process \cite{Sajid2020}. For Ti$_{\text{VV}}$, Fig.~\ref{fig:Ti-VV_in_hBN}a displays the deep-lying localized defect states, indicating room-temperature operation. Connecting to the multiplet electronic structure in Fig.~\ref{fig:Ti-VV_in_hBN}b, the lower $\ket{\pm}$ state is $\ket{g}$ whereas the upper $\ket{\pm}$ state serves as $\ket{e}$. A state in a singlet configuration is used for $\ket{s}$. This definition is also applied to $\text{Mo}_\text{VV}$ as shown in Figs.~\ref{fig:Ti-VV_in_hBN}c and \ref{fig:Ti-VV_in_hBN}. For more candidates of hBN defects suitable for quantum memory, Ref. \cite{cholsuk2023identifying} has recently investigated a large set and identified inherited electric transitions for quantum memory in these defects.\\
\begin{figure*}[ht!]
    \begin{subfigure}[b]{.49\textwidth}
        \includegraphics[width = \textwidth]{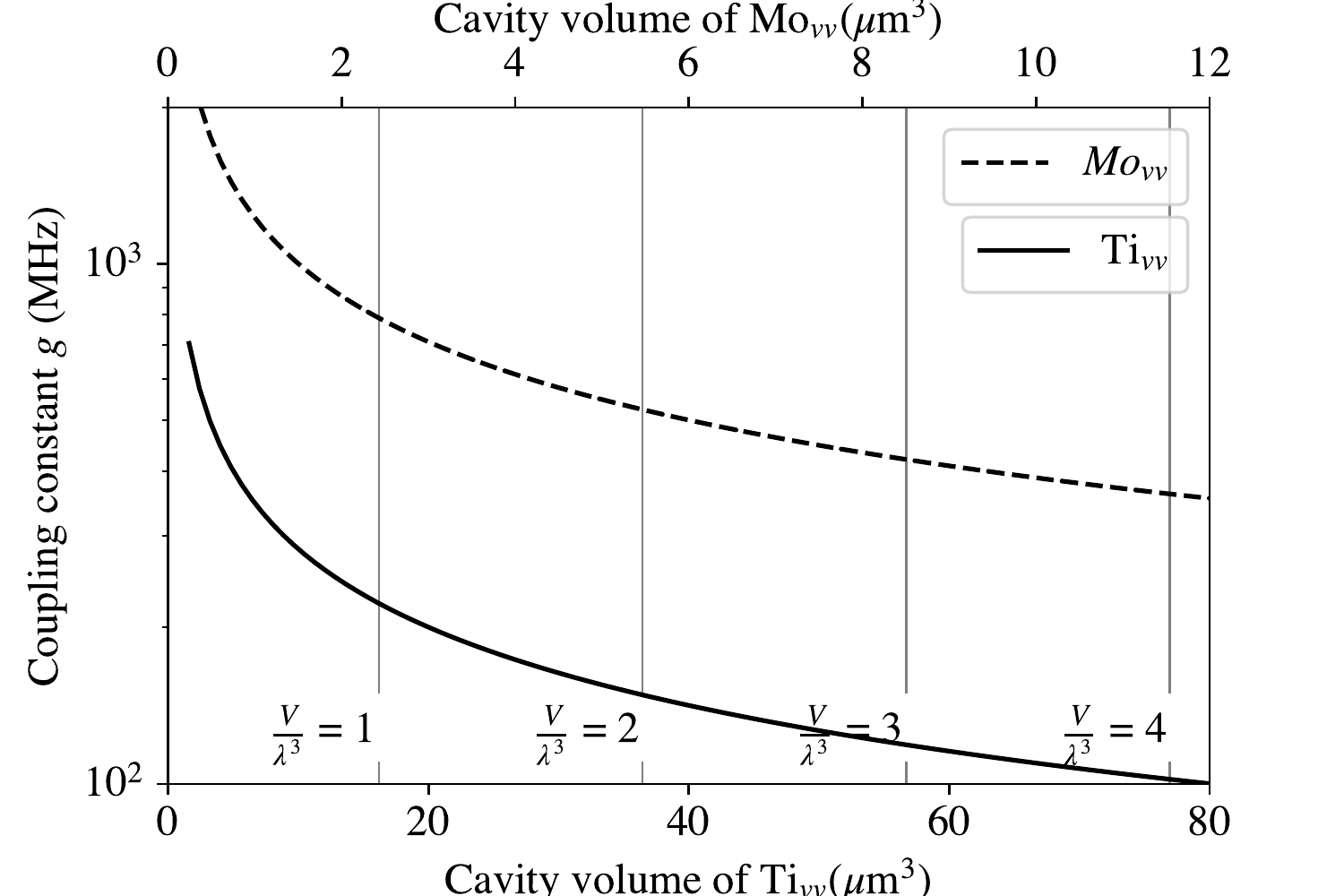}
        \caption{}
        \label{fig:Ti-VV-coupling-constant-volume}
    \end{subfigure}
    \hfill
    \begin{subfigure}[b]{.49\textwidth}
        \includegraphics[width = \textwidth]{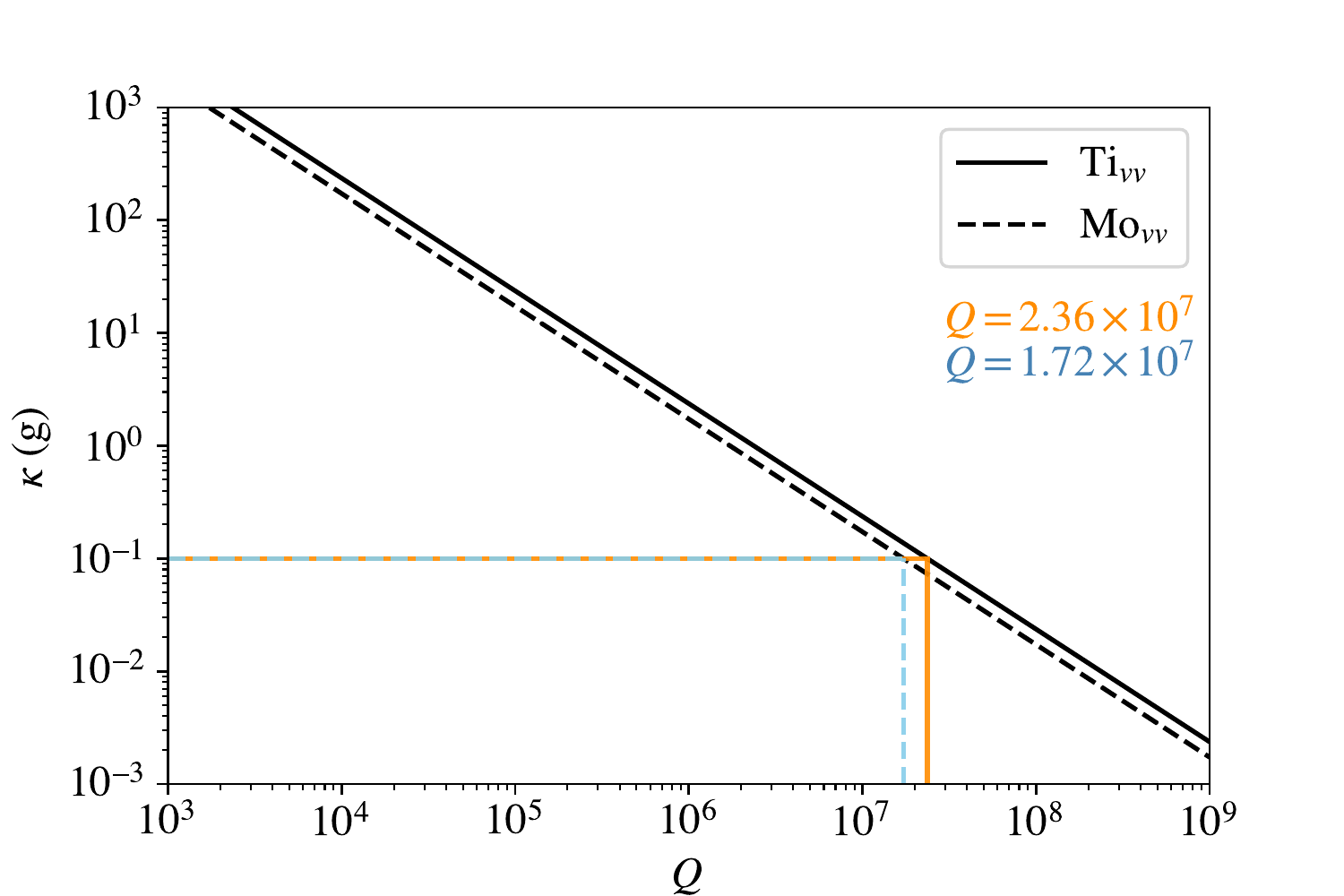}
        \caption{}
        \label{fig:Ti-VV-quality-decay}
    \end{subfigure}
    \caption{ (a) Cavity volume and coupling constant for Ti$_\text{VV}$ (solid line) and Mo$_\text{VV}$ (dashed line). The vertical lines show cavity volume in the unit of the cube of resonance wavelength. (b) Cavity quality factor in a typical range \cite{Fait2021} and decay rate in a unit of $g$ for Ti$_\text{VV}$ and Mo$_\text{VV}$. The annotated blue and orange lines highlight the quality factor resulting in the decay rate one-tenth of the coupling constant $g$.}
\end{figure*}
\indent For any of these defects to be useful as a quantum memory, the defect has to be placed in an optical cavity, which plays two roles. First, the cavity volume determines the coupling constant $g$ between the incoming photon and the transition between $\ket{g}$ and $\ket{e}$ as shown in Eq.~\eqref{eq:couplingConstant-g}. Second, the cavity field decay rate determines the maximum characteristic time $T$ that the writing process can take before losing a photon. In Fig.~\ref{fig:Ti-VV-coupling-constant-volume}, we calculated the coupling constant of the cavity for both defects as a function of cavity volume $V = n \lambda^3$ from $n = 0.1$ to $n=5$. By assuming a resonant transition, the corresponding frequencies $\omega_\text{ge}$ and $\omega_\text{se}$ yielded 744.6 (1518.9) THz and 288.4 (1139.2) THz for Ti$_{\text{VV}}$ (Mo$_{\text{VV}}$), respectively. For the transition dipole moment, we took the values reported in Ref.~\cite{Smart2021} into account. We found that the correlation between cavity volume and the cube of wavelength is preserved \cite{Fait2021}. For the sake of demonstration, we chose $n = 2$ for further study. Here, the coupling constant $g$ at perfect resonance $\Delta=0$ is obtained as
\begin{align}
    g &
    \approx 160 \text{ MHz}  \text{   for Ti}_\text{VV} \\
    g&
    \approx 556 \text{ MHz}  \text{   for Mo}_\text{VV}. 
\end{align}
\indent Suppose that the peak Rabi frequency $\Omega_0 = 100\times g$ and the maximum characteristic time $T = 10/g$ for both defects. In summary,
\begin{align}
    &\Omega = 16.0 \text{ GHz} &T = 62.5 \text{ ns} \text{   for Ti}_\text{VV}\\
    &\Omega = 55.6 \text{ GHz} &T = 18.0 \text{ ns} \text{   for Mo}_\text{VV}. 
\end{align}
These cause HWHM $\sigma_\Delta \approx 10$ (as seen in Fig.~\ref{fig:2-photon-absorption-spectrum-detuning-15}), so the bandwidths according to Eq.~\eqref{eq:Bandwidth_quadratic_formula} yield
\begin{align}
    \Delta &= 1.6 \text{ GHz} \text{   for Ti}_\text{VV} \\
    \Delta &= 5.6 \text{ GHz} \text{   for Mo}_\text{VV}.
\end{align}
\indent With these bandwidths, the cavity of each defect has to store a photon long enough for the memory to map a photonic state to an atomic state using characteristic time $T > 10g$. As shown in Fig.~\ref{fig:cavity-field-decay-population}, the cavity field decay rate typically has $\kappa < 0.01g$. The corresponding quality factor $Q$ is then given by $Q > 2.36\times10^8$  for Ti$_\text{VV}$ and $Q > 1.78\times10^8$ for Mo$_\text{VV}$ as shown in Fig.~\ref{fig:Ti-VV-quality-decay}. From this result, these quality factors are determined directly from the material parameters. The question arises of how we can manipulate the control parameters to decrease the quality factor whereas achieve the writing/reading efficiency greater than 95\%.\\
\indent Here, we propose the theoretical solution based on the fact that the cavity field decay rate is a determining factor of the upper bound of the characteristic time $T$. If we reduce the bandwidth, the required quality factor value can be subsequently lower. From Fig.~\ref{fig:contour-efficiency-writing}, we found that 95\% writing and reading efficiencies can be achieved if we can control the characteristic time $T$ to be greater than $2/g$. With this $T$ constraint, the required quality factor will be reduced to $Q > 2.36\times10^7$ for Ti$_\text{VV}$ and $Q > 1.78\times10^7$ for Mo$_\text{VV}$ as exhibited in Fig.~\ref{fig:Ti-VV-quality-decay} and Eq.~\eqref{eq:decay-quality}.\\
\indent Overall, we achieved to find the scenario enhancing the writing/reading efficiency of hBN to be greater than 95$\%$ while still requiring a cavity within a typical quantity factor range ($10^7$). These values are currently beyond reach for dielectric cavities \cite{Vogl2019} but can be achieved with photonic crystal cavities \cite{pcc}. Comparing the defect candidates,  Mo$_\text{VV}$ requires a lower quality factor from the cavity, but it also requires a smaller cavity volume than Ti$_\text{VV}$. The difference in the quality factor requirement is within the same order of magnitude, while the difference in cavity volume is one order of magnitude apart (a few cubic micrometers and tens of cubic micrometers). 

\section{Conclusion and outlook}
\label{sec:conclusion}
This work has demonstrated how the performance of quantum memories based on an adiabatic mapping process depends on physical and engineering parameters. Those physical parameters include energy levels, decay rates of the three-level structure, and decay rate of the cavity field. We found that all electronic decay rates except the dephasing of the excited state are relevant to the performance of quantum memory. For the engineering parameters, we investigated the characteristic time $T$ and the peak of controlled Rabi frequency $\Omega_0$. These two parameters indicate how quickly a quantum state of light can be mapped to a quantum state of matter. We found that $T$ generally improve mapping efficiency, i.e., the slower the mapping process, the higher the efficiency. However, the cavity field decay rates limit how slow the process can be achieved.\\
\indent We then apply the relation of control parameters to analyze the performance of quantum memory from Ti$_\text{VV}$ and Mo$_\text{VV}$ defects hosted by hBN. We proposed the theoretical viewpoints that both defects can achieve high writing and reading efficiencies ($>95\%$) with a practical range of the optical cavity's quality factor ($Q\approx10^7$). In addition, we found that Ti$_\text{VV}$ and Mo$_\text{VV}$ have the electronic structures that can store a photon at bandwidth in the order of $100$ MHz. However, the promising defect candidates are not limited by these two defects only. We hypothesize that a defect in hBN featuring the intersystem-crossing channel and long lifetime is likely to be a good candidate for a quantum memory. This is being comprehensively studied in our ongoing work.\\
\indent Finally, our performance analysis evaluated between the material's physical parameters and engineering parameters can be further applied to other materials beyond hBN and also other $\Lambda$-structure schemes. This paves the way for theoretically designing and tailoring quantum materials for use in quantum memories. Our work has therefore implications for future long-distance repeater-assisted quantum communication networks and distributed quantum computing architectures.

\section*{Data availability}
All data from this work is available from the authors upon reasonable request.

\section*{Notes}
The authors declare no competing financial interest.

\begin{acknowledgments}
This work is funded by Mahidol University (Fundamental Fund FF-093/2567: fiscal year 2024 by National Science Research and Innovation Fund (NSRF)) and from the NSRF via the Program Management Unit for Human Resources \& Institutional Development, Research and Innovation (grant number B05F650024) and the Deutsche Forschungsgemeinschaft (DFG, German Research Foundation) - Projektnummer 445275953. The authors acknowledge support by the German Space Agency DLR with funds provided by the Federal Ministry for Economic Affairs and Climate Action BMWK under grant number 50WM2165 (QUICK3) and 50RP2200 (QuVeKS). T.V. is funded by the Federal Ministry of Education and Research (BMBF) under grant number 13N16292. T.N. acknowledges financial supports from Mahidol University and National Science and Technology Development Agency (NSTDA). C.C. is grateful to the Development and Promotion of Science and Technology Talents Project (DPST) scholarship by the Royal Thai Government.
\end{acknowledgments}

\bibliographystyle{apsrev}
\bibliography{paper}

\clearpage
\begin{appendix}
\section{Adiabaticity}
\label{sec:adiabatic-appendix}
As this work is scoped on a quantum memory in which a quantum state of light is mapped to an atomic state via adiabatic passage, this section analyzes control parameters and their relation based on that constraint.\\
\subsection{Adiabatic basis}
\label{sec:adiabatic-basis}
\begin{figure}
    \centering
    \includegraphics[width=.5\textwidth]{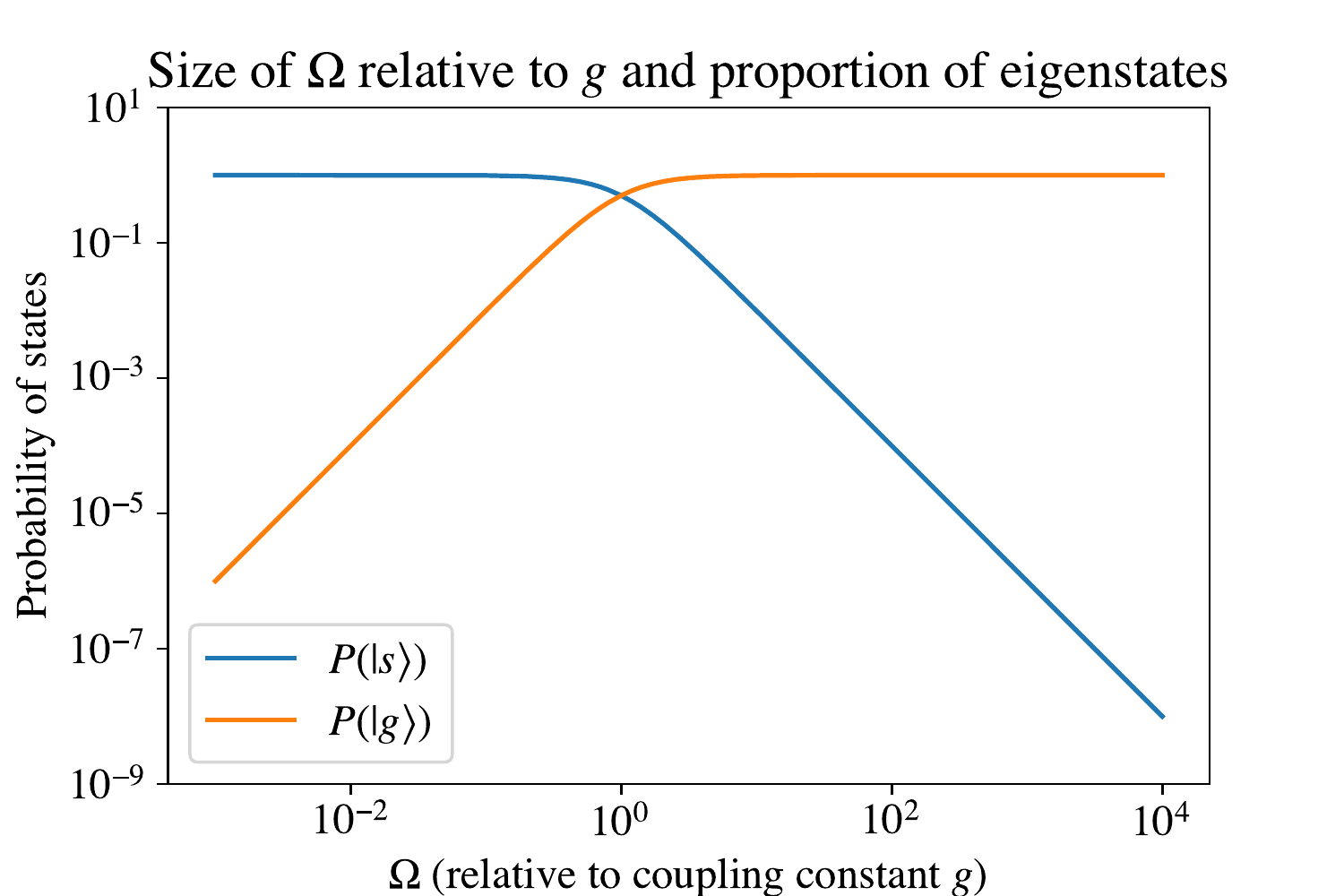}
    \caption{Strength of peak Rabi frequency $\Omega_0$ relative to $g$ and resulting mixing ratio between ground and metastable states.}
    \label{fig:OmegaMixingRatio}
\end{figure}

The adiabatic basis is a basis set formed by the instantaneous eigenvectors of the interacting Hamiltonian in Eq.~\eqref{eq:Hamiltonian}. If the one-photon detuning $\Delta$ is zero, the basis is composed of 
\begin{align}
    \ket{\Phi_0} &= \cos\theta\ket{g} - \sin\theta\ket{s} 
    \label{eq:phi0}, \\
    \ket{\Phi_\pm} &= \frac{\sin\theta}{\sqrt{2}}\ket{g} + \frac{\cos\theta}{\sqrt{2}}\ket{s} \pm \frac{1}{\sqrt{2}}\ket{e} ,
\end{align}
where $\theta = \arctan\frac{g\sqrt{N}}{\Omega}$ is a mixing angle. If the detuning $\Delta$ is not zero, the dark state $\ket{\Phi_0}$ remains the same, but the bright states $\ket{\Phi_+}$ and $\ket{\Phi_-}$  change into 
\begin{align}
    \ket{\Phi_+} &= \sin\theta\cos\phi\ket{g} + \cos\theta\cos\phi\ket{s} - \sin\phi\ket{e}\label{eq:phi_p-non_zero_Delta}\\
    \ket{\Phi_-} &= \sin\theta\sin\phi\ket{g} + \cos\theta\sin\phi\ket{s} + \cos\phi\ket{e},
    \label{eq:phi_m-non_zero_Delta}
\end{align}
where $\arctan 2\phi = \arctan\frac{\sqrt{g^2 + \Omega^2}}{\Delta}$. Note that if $\Delta =0$, this will recover the bright states from a perfect resonance case. Moreover, if $\Delta$ is large enough, then a bright state $|\Phi_+\rangle$ has a very small component overlapping the excited state.\\
\indent The wavefunction will evolve adiabatically if the change in the Hamiltonian is sufficiently slowly. A quantum state of light is mapped into an atomic state by slowly changing the mixing angle $\theta$. Initially, when $\theta = 0$, every atom is in the ground state and couples with the photon number state $\ket{1}$, and slowly changes as $\theta \rightarrow \frac{\pi}{2}$, which results in the $\ket{g,1} \rightarrow \ket{s,0}$ transition. By the adiabatic theorem, if the process is slow enough, it can be guaranteed that the system remains in the dark adiabatic basis. However, the opposite is also possible, in which case, a transition is through the bright states $\ket{\Phi_\pm}$. In contrast to going through a dark state, the mixing angle starts with $\theta = \frac{\pi}{2}$ if the atom is initially in the ground state.\\ 
\indent To change the mixing angle, the Rabi frequency $\Omega(t)$ is changed because the coupling constant $g$ is kept constant during the whole operation. Ideally, since $\theta=0$ occurs when $\omega>>g$, the control field of the Rabi frequency has to be infinitely strong, which is not physically realizable. The electric field has to take a finite value. To answer the question of how strong the Rabi frequency should be, Fig.~\ref{fig:OmegaMixingRatio} is plotted to show the mixing proportion between the ground and metastable states, and the strength of the control field relative to the strength of $g$. It suggests that varying the strength of the control field in the order of $10^{\pm 2}$ relative to the coupling constant $g$ results in the variation of $\ket{g}$ and $\ket{s}$ probabilities in the order of $10^{-4}$.

\subsection{Non-adiabatic correction from detuning}
\label{sec:non-adiabatic-correction}
\begin{figure}[!h]
    \centering
    \includegraphics[width = .5\textwidth]{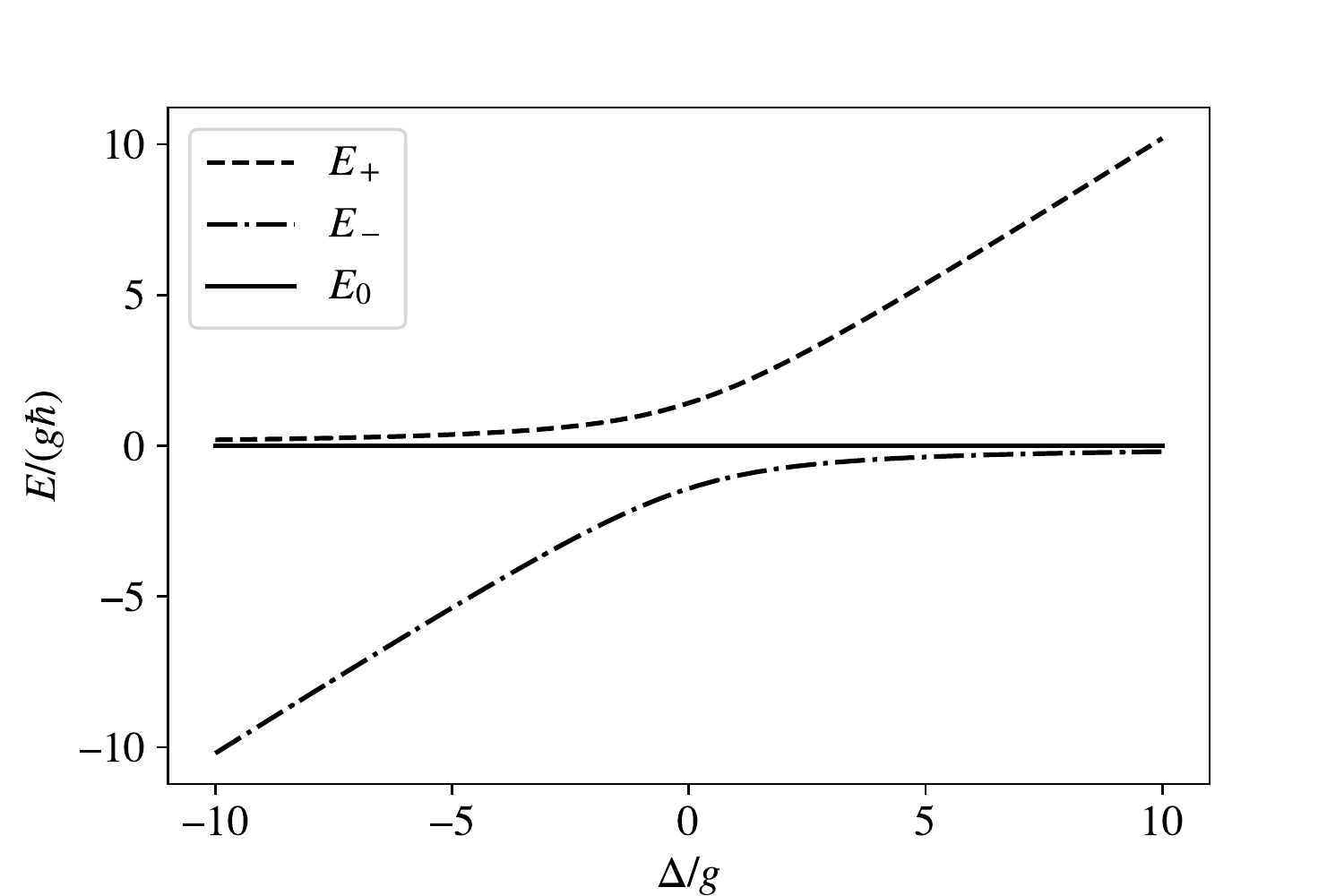}
    \caption{Eigenvalues of adiabatic basis dependence on detuning.}
    \label{fig:Adiabatic_energies-with_Delta-Omega1}
\end{figure}

One-photon detuning of the signal and control field leads to non-adiabaticity because eigenenergies of the adiabatic basis depend on the detuning as visualized in Fig.~\ref{fig:Adiabatic_energies-with_Delta-Omega1}. The eigenenergies are as follows
\begin{align}
    E_0 &= 0\\
    E_\pm &= \frac{1}{2} \left(\Delta \pm \sqrt{\Delta ^2+4 g^2+4 \Omega ^2}\right)
\end{align}
As the detuning grows larger, the difference between $\ket{\Phi_+}$ and $\ket{\Phi_0}$ is narrower. Hence, the adiabatic approximation of population dynamics in these two states is no longer valid.
\onecolumngrid
\section{Expression of projection operators in adiabatic basis}
\label{sec:Projection-in-adiabatic}
 To consider a population decay from an adiabatic state $\ket{j}$ to the state $\ket{i}$, the projection $\ketbra{i}{j}$ in the basis $\qty{\ket{\Phi_{0,\pm}}}$ has to be determined. The expressions of the projection operators $\{\ketbra{i}{j}\}$ in the basis of $\{\ket{\Phi_0},\ket{\Phi_-}, \ket{\Phi_+}\}$ in the respective order are given below. The orthogonal projection $\ketbra{i}{i}$ represents ``pure dephasing''.
\begin{align}
    \ketbra{g}{g} &= \left(
        \begin{array}{ccc}
         \cos ^2 \theta  & \sin \theta \cos \theta  \cos \phi  & \sin \theta  \cos \theta \sin \phi  \\
         \sin\theta \cos\theta \cos\phi & \sin ^2\theta  \cos ^2 \phi & \sin ^2\theta \sin\phi \cos\phi  \\
         \sin\theta \cos\theta \sin\phi & \sin ^2\theta \sin\phi\cos\phi & \sin ^2 \theta \sin ^2 \phi\\
        \end{array}
        \right)\label{eq:g-dephasing}\\
    \ketbra{s}{s} &= \left(
        \begin{array}{ccc}
         \sin ^2\theta & -\sin \theta \cos \theta \sin \phi & -\sin \theta \cos \theta \cos \phi \\
         -\sin \theta \cos \theta \sin \phi & \cos ^2\theta \sin ^2\phi & \cos ^2\theta \sin \phi \cos \phi \\
         -\sin \theta \cos \theta \cos \phi & \cos ^2\theta \sin \phi \cos \phi & \cos ^2\theta \cos ^2\phi \\
        \end{array}
        \right)\label{eq:s-dephasing} \\
    \ketbra{e}{e} &= \left(
        \begin{array}{ccc}
         0 & 0 & 0 \\
         0 & \cos ^2\phi & -\sin \phi \cos \phi \\
         0 & -\sin \phi \cos \phi & \sin ^2\phi \\
        \end{array}
        \right)\\
    \ketbra{g}{s} &= \left(
        \begin{array}{ccc}
         -\sin \theta \cos \theta & \cos ^2\theta \sin \phi & \cos ^2\theta \cos \phi \\
         -\sin ^2\theta \cos \phi & \sin \theta \cos \theta \sin \phi \cos \phi & \sin \theta \cos \theta \cos ^2\phi \\
         -\sin ^2\theta \sin \phi & \sin \theta \cos \theta \sin ^2\phi & \sin \theta \cos \theta \sin \phi \cos \phi \\
        \end{array}
        \right) \\
    \ketbra{g}{e} &= \left(
        \begin{array}{ccc}
         0 & \cos \theta \cos \phi & -\cos \theta \sin \phi \\
         0 & \sin \theta \cos ^2\phi & - \sin \theta \sin \phi \cos \phi \\
         0 & \sin \theta \sin \phi \cos \phi & -\sin \theta \sin ^2\phi \\
        \end{array}
        \right)\\
    \ketbra{s}{e} &= \left(
        \begin{array}{ccc}
         0 & -\sin \theta \cos \phi & \sin \theta \sin \phi \\
         0 & \cos \theta \sin \phi \cos \phi & -\cos \theta \sin ^2\phi \\
         0 & \cos \theta \cos ^2\phi & -\cos \theta \sin \phi \cos \phi \\
        \end{array}
        \right)
\end{align}
\indent Their Hermitian conjugates are straightforwardly obtained by taking the transpose.
\end{appendix}

\end{document}